\documentclass[fleqn,usenatbib]{mnras}

\usepackage{newtxtext,newtxmath}
\usepackage{amsmath}

\usepackage[T1]{fontenc}

\DeclareRobustCommand{\VAN}[3]{#2}
\let\VANthebibliography\thebibliography
\def\thebibliography{\DeclareRobustCommand{\VAN}[3]{##3}\VANthebibliography}

\usepackage{graphicx}	
\usepackage{amsmath}	
\usepackage{enumitem}
\usepackage{tikz}
\usepackage{xcolor}

\newcommand{\betadm}{\beta_\mathrm{dm}}
\newcommand{\pdm}{p_\mathrm{dm}}
\newcommand{\qdm}{q_\mathrm{dm}}
\newcommand{\betaan}{\beta_\mathrm{an}}

\defcitealias{han2022}{H22}

\definecolor{sbblue}{HTML}{4C72B0}
\definecolor{sborange}{HTML}{DD8452}
\definecolor{sbgreen}{HTML}{55A868}

\title[Geometry of the Milky Way’s dark matter halo]{Geometry of the Milky Way’s dark matter from dynamical models of the tilted stellar halo}

\author[A. M. Dillamore et al.]{
Adam M. Dillamore\thanks{E-mail: a.dillamore@ucl.ac.uk (AMD)}
and Jason L. Sanders
\\
Department of Physics and Astronomy, University College London, London, WC1E 6BT, UK\\
}

\date{Accepted XXX. Received YYY; in original form ZZZ}

\pubyear{\the\year{}}

\begin{document}
\label{firstpage}
\pagerange{\pageref{firstpage}--\pageref{lastpage}}
\maketitle

\begin{abstract}
The shape and orientation of the Milky Way's dark matter halo remain poorly constrained. Observations of the accreted stellar halo show that it is triaxial and tilted with respect to the disc. If this configuration is long-lived, it can place constraints on the shape and orientation of the dark matter halo that can support it close to steady state. We use a novel method to fit equilibrium orbit-superposition (Schwarzschild) models to the stellar halo in a realistic Milky Way potential with a tilted dark matter halo. We assume that the long axes of each halo and the disc normal are coplanar. These models are matched to parametric density fits and velocity anisotropy measurements of \textit{Gaia} Sausage-Enceladus (GSE) stars at radii $r\in[6,60]$~kpc. The observations are consistent with a (near-)prolate dark matter halo whose density has a short-to-long axis ratio of $\qdm=0.87_{-0.09}^{+0.05}$. The long axis is inclined at an angle of $\betadm=43_{-8}^{+22}\,^\circ$ to the disc plane, which exceeds the stellar halo tilt by $\approx18^\circ$. Spherical haloes cannot support the observed structure of the GSE in equilibrium. The best-fitting dynamical GSE model has a radius-dependent shape and orientation; between radii of 6 and 60~kpc the tilt increases from $\beta_*(r)\approx10^\circ$ to $\approx35^\circ$. Our model provides a good fit to the observed triaxial structure and dynamics of the GSE. It is therefore an excellent source of realistic initial conditions for simulations of the halo, such as for investigating perturbations from satellites or the Galactic bar.
\end{abstract}




\section{Introduction}

\textit{Gaia} \citep{gaia} ushered in a revolution in the study of the Milky Way's stellar halo. Kinematic and chemical data revealed that it is dominated by a population of relatively metal-rich stars ([Fe/H]~$>-2$) on highly eccentric orbits \citep{belokurov2018,helmi2018}. This likely originated in a single merger event 8-11 Gyr ago \citep{belokurov2020,bonaca2020} with a dwarf galaxy of stellar mass $\sim3\times10^8\mathrm{M}_\odot$ \citep[e.g.][]{mackereth2020}, named the \textit{Gaia} Sausage-Enceladus (GSE). Various studies have measured its velocity anisotropy profile \citep[e.g.][]{lancaster2019,bird2021,iorio2021,chandra2025}, finding that it remains highly radially anisotropic at least as far out as $20$~kpc from the Galactic centre.

Attempts to model the dynamics of the stellar halo usually assume that it is axisymmetric \citep[e.g.][]{das2016,hattori2021}. However, observations from
\textit{Gaia} suggest that its structure is more complex. \citet{iorio2018} and \cite{iorio2019} used RR Lyrae stars to show that it is triaxial, with the major axis lying in the Galactic plane at an angle of $\sim70^\circ$ to the Sun-Galactic centre line. The shape is radius-dependent, with the halo becoming more triaxial with increasing radius. Their best-fitting model is also tilted out of the Galactic plane by $\beta=20^\circ$. The major axis is approximately aligned with two previously known overdensities of stars, the Hercules-Aquila Cloud \citep[HAC;][]{belokurov2007} and the Virgo Overdensity \citep[VOD;][]{vivas2001,juric2008}. In fact these structures have been shown to be both dynamically \citep{simion2019,balbinot2021} and chemically \citep{perottoni2022,ye2024} indistinguishable from the GSE itself. They therefore provide evidence that the GSE debris remains in an unmixed, triaxial, and tilted configuration. \citet{han2022}, henceforth \citetalias{han2022}, fitted a triaxial density model to GSE stars from the H3 survey \citep{conroy2019}. They found that the data are fitted well by a near-prolate model with its long axis tilted by $25^\circ$ above the Galactic plane. The density profile is a doubly-broken power law with break radii at $r=12$ and 28~kpc. \citet{lane2023} similarly found that the GSE remnant is nearly prolate and tilted, albeit with a lower inclination of $16^\circ$.

The non-axisymmetric and tilted structure of the GSE debris has led some authors to propose that it results from a much more recent merger event \citep{donlon2019,donlon2023}. Indeed, in an axisymmetric potential such a configuration mixes and loses its tilt in less than $\sim1$~Gyr \citep{han2022b}. However, estimates of stellar ages for the GSE and other halo stars indicate that the merger was complete by $\sim8$ Gyr ago \citep{dimatteo2019,gallart2019,belokurov2020,bonaca2020,horta2024_gse}. Instead, the observed structure may be supported in a non-axisymmetric potential without mixing. \citet{naidu2021} ran $N$-body simulations of the GSE merger, and successfully reproduced a triaxial structure with axis ratios $10:7.9:4.5$ and a tilt of $\sim35^\circ$. \citet{han2022b} showed that this configuration can only persist until the present day if the Milky Way's dark matter halo is non-spherical and tilted with respect to the disc. This is consistent with theoretical expectations; triaxial density distributions tend to be supported by box orbits aligned with their major axes \citep{binney_tremaine}, which require the potential to be elongated roughly along the same direction. Such orbits do not exist in axisymmetric potentials. \citet{han2023a,han2023} found that tilted stellar and dark matter haloes are common and long-lasting in cosmological simulations, and that they can warp Galactic discs. Similarly \citet{dillamore2022} showed that tilted dark matter haloes and slowly tilting discs persist for many Gyr after GSE-like mergers. Hence as a first approximation it is reasonable to model the GSE debris as being close to equilibrium in a potential with a tilted dark matter halo.

It is clear that the geometry and symmetry of a steady-state stellar halo are directly related to those of the Galactic potential, and therefore the dark matter halo \citep{an2017,han2023}. There is still very little consensus on the shape or orientation of our Galaxy's dark matter halo \citep{hunt2025}. Most previous constraints on the shape have come from modelling of stellar streams \citep[e.g.][]{law2010,koposov2010,bowden2014,kupper2015,bovy2016,malhan2019,erkal2019,koposov2023,ibata2024}. The majority of these results are consistent with a spherical dark matter halo \citep{hunt2025}, although the Orphan-Chenab stream favours either a prolate or oblate halo at $r\gtrsim15$~kpc \citep{erkal2019,koposov2023}. \citet{vasiliev_tango} modelled the Sagittarius stream including perturbations from the Large Magellanic Cloud (LMC). They found evidence that the halo is oblate and aligned with the Galactic plane at small radii but nearly perpendicular to the plane at $r\gtrsim60$~kpc. \citet{nibauer2025} measured the acceleration field of the Milky Way from the 6D track of the GD-1 stream. They favoured a triaxial dark matter halo with density axis ratios $1:0.75:0.7$, with the long axis tilted by $18^\circ$ above the Galactic plane. \citet{dodd2022} and \citet{woudenberg2024} used the effects of resonances in non-spherical potentials \citep{yavetz2021,yavetz2023} on the Helmi streams to precisely constrain the halo's shape, assuming that it is oblate (with axis ratios $1.2:1$) and triaxial ($1.204:1.013:1$) respectively.

In this study we adopt a novel approach to constrain the geometry of the dark matter halo: dynamical modelling of the tilted triaxial stellar halo. Various modelling approaches are available for fitting a stellar distribution. In an axisymmetric system, action-based distribution function (DF) models can be used to fit the 6D phase space distribution of stars. If the actions $\boldsymbol{J}$ \citep[estimated using the St\"{a}ckel fudge;][]{binney2012} are integrals of motion, then the DF automatically corresponds to a steady-state distribution of stars. This approach was used by \citet{hattori2021} to constrain the flattening of the Milky Way's dark matter halo, and various studies have applied it to the globular cluster population \citep[e.g.][]{posti2019,wang2022}. However, this approach is unsuitable for our purposes, since potentials with tilted triaxial dark matter haloes do not possess enough integrals of motion \citep[though for methods that may be applicable in some triaxial potentials see][]{vandeven2003,sanders2014,sanders2015}.

A more appropriate technique in triaxial potentials is Schwarzschild modelling \citep{schwarzschild1979}. This involves postulating a potential and integrating a set of orbits over many periods to generate an \textit{orbit library}. The weights of each orbit are then optimized to find the best superposition that matches the target density. Additional constraints from kinematics (e.g. velocity anisotropy) can also be included. This process can then be repeated with different postulated potentials to find the best fit to the data. Since this method does not require any particular symmetry \citep[though for a discussion of chaos in triaxial potentials see][] {merritt1996}, it is ideal for fitting non-axisymmetric systems \citep[e.g.][]{schwarzschild1979,schwarzschild1982}. \citet{ling2025} recently demonstrated the power of this method to model the stellar halo. We therefore select Schwarzschild modelling to fit the density of the GSE as reported by \citetalias{han2022}. This will place constraints on the tilt and axis ratios of the dark matter halo required to support the observed configuration in equilibrium. The results can be compared with Milky Way analogues in cosmological simulations as a test of the $\Lambda$CDM paradigm of cosmology.

The rest of this paper is arranged as follows. In Section~\ref{section:data} we outline the observations of the stellar halo that we aim to fit. We describe the modelling procedure in Section~\ref{section:schwarzschild} and present the results in Section~\ref{section:results}. We discuss our findings further in Section~\ref{section:discussion} and finally summarise our conclusions in Section~\ref{section:conclusions}.

\begin{figure*}\label{fig:coords}
    \centering
    \begin{minipage}{0.3\textwidth}
        \centering
        \begin{tikzpicture}[scale=1.2,>=stealth]
            \def\alphahalo{204.33} 
            \def\r{2}
            
            \draw[->, very thick] (-\r,0) -- (\r,0) node[anchor=west] {\Large $x$};
            \draw[->, very thick] (0,-\r) -- (0,\r) node[anchor=south] {\Large $y$};
            
            \draw[->, very thick, sbblue] ({-\r*cos(\alphahalo)}, {-\r*sin(\alphahalo)}) -- ({\r*cos(\alphahalo)}, {\r*sin(\alphahalo)}) node[anchor=north east] {\Large $x'$};
            \draw[->, very thick, sbblue] ({\r*sin(\alphahalo)}, {-\r*cos(\alphahalo)}) -- ({-\r*sin(\alphahalo)}, {\r*cos(\alphahalo)}) node[anchor=north west] {\Large $y'$};
            
            \begin{scope}[rotate=\alphahalo]
            \draw[black, thick] (0,0) ellipse (1.5 and 1.5*0.81);
            \end{scope}
            
            \draw[->, very thick, gray] (0.5,0) arc (0:\alphahalo:0.5);
            \node at (0.4,0.55) {\Large $\alpha_*$};
            
            \filldraw[black] (-1,0) circle (1.5pt);
            \draw[black, very thick] (-1,0) circle (4pt);
            
            \node[anchor=west] at (-2,2.5) {\Large Top-down};

            \node[anchor=south west] at (0.1,1.3) {\Large Stellar halo};  
        
        \end{tikzpicture}
    
    \end{minipage}
    \hfill
    \begin{minipage}{0.3\textwidth}
    
        \begin{tikzpicture}[scale=1.2,>=stealth]
            \def\thetastar{25.39} 
            \def\thetadm{40}
            \def\r{2}
            
            \draw[->, very thick, sbblue] (-\r,0) -- (\r,0) node[anchor=west] {\Large $x'$};
            \draw[->, very thick] (0,-\r) -- (0,\r) node[anchor=south] {\Large $z$};
            
            \draw[->, very thick, sborange] ({-\r*cos(\thetastar)}, {-\r*sin(\thetastar)}) -- ({\r*cos(\thetastar)}, {\r*sin(\thetastar)}) node[anchor=south west] {\Large $X_*$};
            \draw[->, very thick, sborange] ({\r*sin(\thetastar)}, {-\r*cos(\thetastar)}) -- ({-\r*sin(\thetastar)}, {\r*cos(\thetastar)}) node[anchor=south east] {\Large $Z_*$};

            
            \begin{scope}[rotate=\thetastar]
            \draw[black, thick] (0,0) ellipse (1.5 and 1.5*0.73);
            \end{scope}
            
            \draw[->, very thick, gray] (0.5,0) arc (0:\thetastar:0.5);
            \node at (0.3,0.35) {\Large $\beta_*$};

            \filldraw[black] ({-cos(204.33)},0) circle (1.5pt);
            \draw[black, very thick] ({-cos(204.33)},0) circle (4pt);

            \node[anchor=west] at (-2,2.5) {\Large Edge-on};

            \node[anchor=south west] at (0.1,1.2) {\Large Stellar halo};            
        \end{tikzpicture}
    \end{minipage}
    \hfill
    \begin{minipage}{0.3\textwidth}
    
        \begin{tikzpicture}[scale=1.2,>=stealth]
            \def\thetastar{25.39}
            \def\thetadm{40} 
            \def\r{2}
            
            \draw[->, very thick, sbblue] (-\r,0) -- (\r,0) node[anchor=west] {\Large $x'$};
            \draw[->, very thick] (0,-\r) -- (0,\r) node[anchor=south] {\Large $z$};

            
            \draw[->, very thick, sbgreen] ({-\r*cos(\thetadm)}, {-\r*sin(\thetadm)}) -- ({\r*cos(\thetadm)}, {\r*sin(\thetadm)}) node[anchor=south west] {\Large $X_\mathrm{dm}$};
            \draw[->, very thick, sbgreen] ({\r*sin(\thetadm)}, {-\r*cos(\thetadm)}) -- ({-\r*sin(\thetadm)}, {\r*cos(\thetadm)}) node[anchor=south east] {\Large $Z_\mathrm{dm}$};
            
            \begin{scope}[rotate=\thetadm]
            \draw[black, dashed, thick] (0,0) ellipse (1.5 and 1.5*0.7);
            \end{scope}
            
            \draw[->, very thick, gray] (0.5,0) arc (0:\thetadm:0.5);
            \node at (0.9,0.35) {\Large $\betadm$};

            \filldraw[black] ({-cos(204.33)},0) circle (1.5pt);
            \draw[black, very thick] ({-cos(204.33)},0) circle (4pt);

            \node[anchor=west] at (-2,2.5) {\Large Edge-on};

            \node[anchor=south west] at (0.05,1.3) {\Large DM halo};            
        \end{tikzpicture}
    \end{minipage}
    
    \caption{Coordinate systems used in this paper. \textbf{Left-hand panel:} Top-down projection (from the North Galactic Pole) showing the $(x,y,z)$ and $(x',y',z)$ coordinates. The $\odot$ symbol marks the position of the Sun, and the ellipse illustrates the orientation of the stellar halo's major axis. \textbf{Middle panel:} Edge-on projection to the Galactic disc (viewed down the $y'$-axis). The orange axes indicate the stellar halo coordinate system $(X_*,Y_*,Z_*)$. \textbf{Right-hand panel:} as above, but showing the dark matter halo coordinate system $(X_\mathrm{dm}, Y_\mathrm{dm}, Z_\mathrm{dm})$ in green. The black dashed ellipse illustrates the orientation of the dark matter halo.}
\end{figure*}
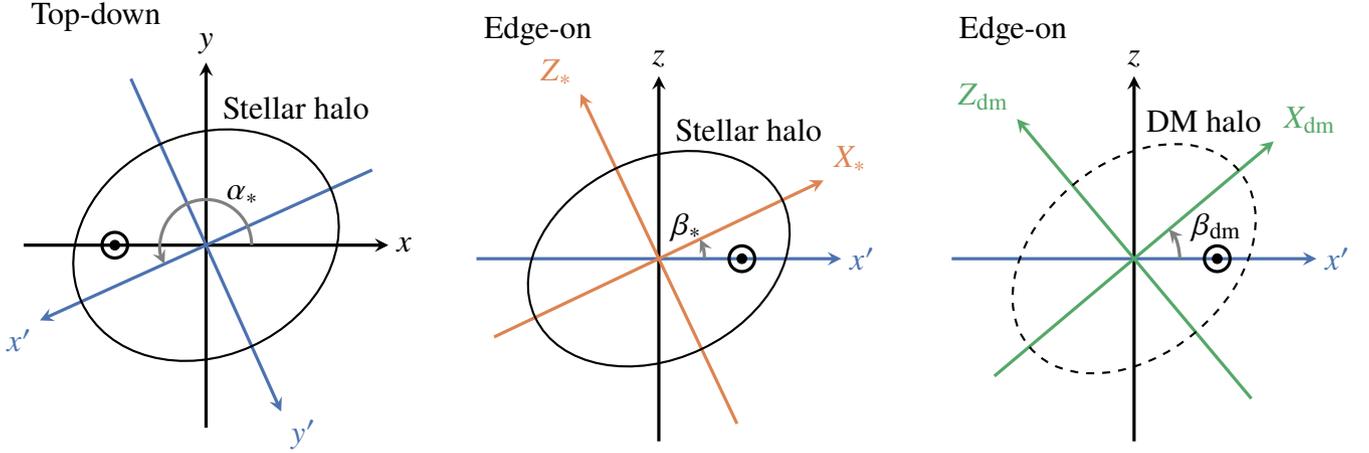

\begin{figure}
  \centering
  \includegraphics[width=\columnwidth]{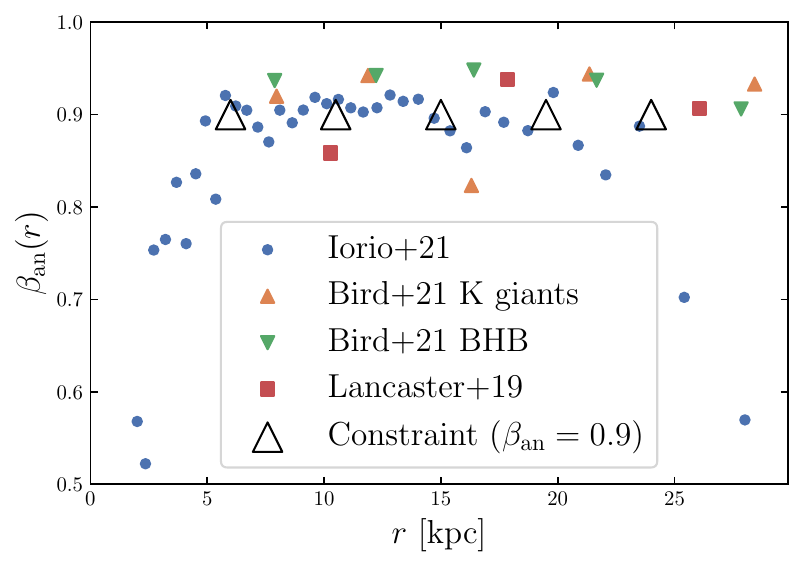}
  \caption{Constraints on the anisotropy profile of GSE stars. The coloured points indicate measurements from RR Lyrae \citep[blue circles;][]{iorio2021}, K giants and BHBs \citep[orange and green triangles;][]{bird2021}, and BHBs from \citet{lancaster2019} (red squares). The black unfilled triangles indicate the constraint chosen for our fit.} 
   \label{fig:beta_an_data}
\end{figure}
\section{Data}\label{section:data}
We use two sets of observational data to fit dynamical models to the stellar halo. These are measurements of the GSE's 3D density distribution and the radial profile of its velocity anisotropy.

\subsection{Stellar halo density}\label{section:data_density}
For the GSE density we use the parametric model fitted to observational data by \citetalias{han2022}. They selected giant stars belonging to the GSE using a cut in the [Fe/H]-[$\alpha$/Fe] plane, and the requirements that $\mathrm{log}\,g<3.5$ and eccentricity $e>0.7$. The density model is ellipsoidally-stratified and tilted with respect to the disc, and has a triple power-law density profile.

Let $(x,y,z)$ be the right-handed Galactocentric coordinate system where the Sun is located at $(-8.2,0,0)$~kpc \citep{bland-hawthorn2016} and Galactic rotational motion is locally in the $+y$ direction. For clearer visualisation we define a new coordinate system $(x',y',z)$ which is related to $(x,y,z)$ by a rotation about the $z$-axis. We choose that the major (and minor) axes of the stellar halo lie in the $x'$-$z$ plane, with the intermediate axis along the $y'$-axis. The coordinates are then related by
\begin{equation}
    \begin{pmatrix}
        x' \\ y' \\ z
    \end{pmatrix}=
    \begin{pmatrix}
        \mathrm{cos}\,\alpha_* & \mathrm{sin}\,\alpha_* & 0\\
        -\mathrm{sin}\,\alpha_* & \mathrm{cos}\,\alpha_* & 0\\
        0 & 0 & 1
    \end{pmatrix}
    \begin{pmatrix}
        x \\ y \\ z    
    \end{pmatrix},
\end{equation}
where $\alpha_*$ is the azimuth of the stellar halo's major axis. The stellar halo density is defined in another coordinate system $(X_*,Y_*,Z_*)$, tilted by an angle $\beta_*$ about the $y'$-axis. This is defined by the transformation
\begin{equation}
    \begin{pmatrix}
        X_* \\ Y_* \\ Z_*
    \end{pmatrix}=
    \begin{pmatrix}
        \mathrm{cos}\,\beta_* & 0 & \mathrm{sin}\,\beta_*\\
        0 & 1 & 0 \\
        -\mathrm{sin}\,\beta* & 0 & \mathrm{cos}\,\beta_*
    \end{pmatrix}
    \begin{pmatrix}
        x' \\ y' \\ z    
    \end{pmatrix},
\end{equation}
where we choose to take $\beta_*>0$ and $0\leq\alpha_*<2\pi$. The relationships between these three coordinate systems are illustrated in the left-hand and middle panels of Fig.~\ref{fig:coords}.

The stellar halo density model is defined in terms of the ellipsoidal radius,
\begin{align}
    r_*&\equiv\sqrt{X_*^2 + (Y_*/p_*)^2 + (Z_*/q_*)^2},
\end{align}
where the axis ratios are in the range $0<p_*,q_*\leq1$. The $X_*$ axis therefore lies along the major axis of the stellar halo, which is inclined at an angle $\beta_*$ to the Galactic plane. \citetalias{han2022} fitted a double-broken power-law density profile to the GSE component of the stellar halo,
\begin{align}\label{eq:rho_dpl}
    \rho_*(r_*)=\rho_{*0}\times\begin{cases}
        (r_*/r_\mathrm{b,1})^{-\gamma_1} & r_* \leq r_\mathrm{b,1} \\
        (r_*/r_\mathrm{b,1})^{-\gamma_2} & r_\mathrm{b,1} < r_* \leq r_\mathrm{b,2} \\
        (r_\mathrm{b,2}/r_\mathrm{b,1})^{-\gamma_2}\,(r_*/r_\mathrm{b,2})^{-\gamma_3} & r_* > r_\mathrm{b,2},
    \end{cases}
\end{align}
where $\rho_{*0}$ is the density normalisation (unimportant for this study), $r_\mathrm{b,1}$ and $r_\mathrm{b,2}$ are the inner and outer break radii, and $\{\gamma_i\}$ are the three power-law slopes. The fitted parameters of the fiducial model and their uncertainties are given in Table~\ref{tab:halo_parameters}. Henceforth we refer to this density model as the `data' for brevity.
\begin{table}
    \centering
    \caption{Parameters of the fiducial stellar halo density fit from \citetalias{han2022}. $M_*$ is the total mass (unimportant for this study); $\alpha_*$ and $\beta_*$ are the Galactic longitude and latitude of the major axis as viewed from the Galactic centre; $p_*$ and $q_*$ are the $Y_*/X_*$ and $Z_*/X_*$ axis ratios (see Fig.~\ref{fig:coords}); $\{r_{\mathrm{b},i}\}$ are the density profile break radii; and $\{\gamma_i\}$ are the power-law slopes. In cases where the upper and lower uncertainties are different we have taken their mean.}
    \begin{tabular}{c|c|c}
    \hline
        Parameter & Value & Uncertainty \\
        \hline
        $\mathrm{log}_{10}(M_*/\mathrm{M}_\odot)$ & 8.76 & 0.02 \\
        $\alpha_*$ [$^\circ$] & 204.33 & 5.23 \\
        $\beta_*$ [$^\circ$] & 25.39 & 3.16 \\
        $p_*$ & 0.81 & 0.03 \\
        $q_*$ & 0.73 & 0.02 \\
        $r_\mathrm{b,1}$ [kpc] & 11.85 & 0.82 \\
        $r_\mathrm{b,2}$ [kpc] & 28.33 & 1.30 \\
        $\gamma_1$ & 1.70 & 0.20 \\
        $\gamma_2$ & 3.09 & 0.11 \\
        $\gamma_3$ & 4.58 & 0.11 \\
        
    \end{tabular}
    \label{tab:halo_parameters}
\end{table}
\subsection{Anisotropy profile}\label{section:data_anisotropy}
We also use kinematic measurements of GSE stars to constrain our models. They are summarised by the anisotropy parameter, defined by
\begin{equation}
    \betaan\equiv1-\frac{1}{2}\frac{\sigma_t^2}{\sigma_r^2},
\end{equation}
where $\sigma_r$ and $\sigma_t$ are the radial and tangential velocity dispersions ($\sigma_t^2=\sigma_\theta^2+\sigma_\phi^2$). The anisotropy of the stellar halo is well-constrained at radii $r\lesssim25$~kpc \citep[e.g.][]{cunningham2019,iorio2021,hattori2021,bird2021,han2024,chandra2025}. It is known to be large ($\betadm\gtrsim0.5$), driven largely by the dominance of GSE stars on highly radial orbits. Since the density model described in Section~\ref{section:data_density} is fit to a selection of GSE stars \citepalias{han2022}, we must draw a distinction between the anisotropy of the stellar halo as a whole and that of the GSE only (the latter being generally higher in value). Various studies have separated the stellar halo population and measured $\betaan(r)$ for the identified GSE component \citep[e.g.][]{lancaster2019,bird2021,iorio2021,lane2025}. Some of these are summarised in Fig.~\ref{fig:beta_an_data}. They agree that the GSE anisotropy is approximately constant at $\betaan\approx0.8-0.95$ at radii in the range $6\lesssim r/\mathrm{kpc} \lesssim24$. We therefore take $\betaan=0.9$ between $r=6$ and 24 kpc (see Section~\ref{section:optimization} for a detailed description of the constraint used in the fit). At larger radii $\betaan$ is more uncertain, so we do not impose constraints on the model kinematics at $r>24$~kpc.

\section{Schwarzschild Modelling}\label{section:schwarzschild}
In this section we detail the procedure used for Schwarzschild modelling of the stellar halo. We describe the potential model in Section~\ref{section:potential}, the orbit library in Section~\ref{section:library}, the model optimization in Section~\ref{section:optimization}, and the goodness of fit estimation in Section~\ref{section:goodnessoffit}. For a review of Schwarzschild modelling methods see \citet{vasiliev2013}.

\subsection{Potential}\label{section:potential}
\subsubsection{Baryonic components}

We keep the baryonic components of the potential fixed throughout this study. These are taken from the \citet{mcmillan17} potential, and consist of two stellar discs, two gas discs, and a bulge. The stellar halo has negligible mass so does not contribute to our model potential.

The two stellar discs represent the thin and thick discs of the Milky Way. Their density distributions have the functional form,
\begin{equation}
    \rho_\mathrm{d}(R,z)=\frac{\Sigma_0}{2z_\mathrm{d}}\,\mathrm{exp}\left(-\frac{|z|}{z_\mathrm{d}}-\frac{R}{R_\mathrm{d}}\right),
\end{equation}
where $\Sigma_0$ is the central surface density, $R_\mathrm{d}$ is the scale radius, and $z_\mathrm{d}$ the scale height.

The gas discs represent HI and molecular gas, and have the density,
\begin{equation}
    \rho_\mathrm{d}(R,z)=\frac{\Sigma_0}{4z_\mathrm{d}}\,\mathrm{exp}\left(-\frac{R_\mathrm{m}}{R}-\frac{R}{R_\mathrm{d}}\right)\,\mathrm{sech}^2\left(\frac{z}{2z_\mathrm{d}}\right),
\end{equation}
where $R_\mathrm{m}$ is the scale length of an inner hole. The values of the parameters for each disc component are given in Table~\ref{tab:disc_parameters}.

\begin{table}
    \centering
    \caption{Parameters of the disc components of the potential used in this study, taken from \citet{mcmillan17}.}
    \begin{tabular}{c|c|c|c|c}
    \hline
        Parameter & Thin disc & Thick disc & HI gas & Molecular gas \\
        \hline
        $\Sigma_0$ [$\mathrm{M}_\odot\,\mathrm{pc}^{-2}$] & 896 & 183 & 53.1 & 2180 \\
        $R_\mathrm{d}$ [kpc] & 2.50 & 3.02 & 7 & 1.5 \\
        $z_\mathrm{d}$ [kpc] & 0.3 & 0.9 & 0.085 & 0.045 \\
        $R_\mathrm{m}$ [kpc] & - & - & 4 & 12 \\
    \end{tabular}
    \label{tab:disc_parameters}
\end{table}

The bulge component has the oblate axisymmetric density profile,
\begin{align}
    \rho_\mathrm{b}(r')&=\frac{\rho_{0,\mathrm{b}}}{(1+r'/r_0)^\alpha}\,\mathrm{exp}\left[-(r'/r_\mathrm{cut})^2\right],\\
    r'&\equiv\sqrt{R^2+(z/q)^2},
\end{align}
where $\alpha=1.8$, $r_0=0.075$~kpc, $r_\mathrm{cut}=2.1$~kpc, and $q=0.5$.

\subsubsection{Dark matter halo}
In this paper we use a rigid analytical model of the dark matter halo (as opposed to fitting a Schwarzschild model to the dark matter itself). Our aim is to constrain the shape and orientation of the dark matter halo based on the stellar halo density. Our triaxial dark halo component is therefore based on the spherical halo in the \citet{mcmillan17} potential, but with variable axis ratios and orientation.

We define a new coordinate system $(X_\mathrm{dm}, Y_\mathrm{dm}, Z_\mathrm{dm})$ aligned with the principal axes of the dark matter halo. In order to match the symmetry of the stellar distribution, we consider dark matter halo models which are tilted in the same plane as the stellar halo density model. The major axes of the two haloes are related by a rotation about the stellar halo's intermediate axes. In other words $(X_\mathrm{dm},Y_\mathrm{dm},Z_\mathrm{dm})$ are related to $(x',y',z)$ by a rotation about the $y'$-axis, so $Y_\mathrm{dm}=Y_*=y'$. Note that the angle $\alpha_*$ is therefore unimportant in this model, and can be assumed to be fixed. We choose to follow the conventions of \citetalias{han2022} and define $\beta_\mathrm{dm}$ to be the tilt of the dark matter halo's major axis relative to the Galactic plane (i.e. the angle between the $x'$ and $X_\mathrm{dm}$ axes). The halo orientation is therefore defined by
\begin{equation}
    \begin{pmatrix}
        X_\mathrm{dm} \\ Y_\mathrm{dm} \\ Z_\mathrm{dm}
    \end{pmatrix}=
    \begin{pmatrix}
        \mathrm{cos}\,\betadm & 0 & \mathrm{sin}\,\betadm\\
        0 & 1 & 0 \\
        -\mathrm{sin}\,\betadm & 0 & \mathrm{cos}\,\betadm
    \end{pmatrix}
    \begin{pmatrix}
        x' \\ y' \\ z    
    \end{pmatrix}.
\end{equation}
This coordinate system is illustrated in the right-hand panel of Fig.~\ref{fig:coords}. Similarly to the stellar halo, the density is expressed in terms of the radial coordinate,
\begin{align}
    r_\mathrm{dm}&\equiv\sqrt{X_\mathrm{dm}^2 + (Y_\mathrm{dm}/p_\mathrm{dm})^2 + (Z_\mathrm{dm}/q_\mathrm{dm})^2}.
\end{align}
We emphasise that throughout this paper all axis ratios (including $\pdm$ and $\qdm$) describe the shape of the halo's density distribution, not its potential. Note that we do not place limits on the relative magnitudes of $\pdm$ and $\qdm$, so the $Y_\mathrm{dm}$ and $Z_\mathrm{dm}$ axes can each be either the minor or intermediate axis.

The halo has a Navarro-Frenk-White \citep[NFW;][]{NFW} profile,
\begin{align}
    \rho_\mathrm{h}(r_\mathrm{dm})&=\frac{\rho_{0,\mathrm{h}}}{\left(r_\mathrm{dm}/r_\mathrm{h}'\right){\left(1+r_\mathrm{dm}/r_\mathrm{h}'\right)^2}},\\
    r_\mathrm{h}'&\equiv (p_\mathrm{dm}q_\mathrm{dm})^{-1/3}\,r_\mathrm{h},
\end{align}
where $\rho_{0,\mathrm{h}}$ is the density normalisation and $r_\mathrm{h}$ the scale radius of the \citet{mcmillan17} halo. The multiplication of the scale radius by $(p_\mathrm{dm}q_\mathrm{dm})^{-1/3}$ ensures that the spherically averaged density profile is approximately conserved when $p_\mathrm{dm}$ and $q_\mathrm{dm}$ vary. The parameters of the \citet{mcmillan17} halo are $\rho_{0,\mathrm{h}}=0.00854~\mathrm{M}_\odot\,\mathrm{pc}^{-3}$ and $r_\mathrm{h}=19.6$~kpc, while we vary $p_\mathrm{dm}$ and $q_\mathrm{dm}$ in our model.

We generate a set of 2D grids of halo potential parameters. We have three parameters $(\betadm,\qdm,\pdm)$ to vary, of which we expect $\betadm$ and $\qdm$ to be most important in matching the shape and orientation of the tilted stellar halo. This is because these most strongly affect the shape of the potential in the plane of tilting $(x',z)$. We therefore focus on $(\betadm,\qdm)$ grids with fixed prescriptions for $\pdm$. In future work we plan to generalise this model to allow arbitrary shapes and orientations. We use the grid $\betadm\in\{0^\circ,2.5^\circ,...,87.5^\circ, 90^\circ\}$ for the tilt. This spans the full range of values in which the major axes of the stellar and dark matter haloes lie in the same quadrant of the $(x',z)$ plane. This is required for the elongation of the stellar halo to be supported by box orbits \citep{binney_tremaine}. The flattening spans the range $\qdm\in\{0.4,0.425,...0.975,1\}$. This approximately matches the $\pm3\sigma$ range of short-to-long axis ratios in Milky Way analogues from the Illustris simulations \citep{chua2019}.

We use two different prescriptions for the intermediate/long axis ratio $\pdm$. Firstly we set $\pdm=\qdm$, which gives prolate haloes symmetric about the $X_\mathrm{dm}$ major axis. This is in line with the expectation from \citet{drakos2019} that GSE-like mergers tend to produce prolate dark matter haloes. Secondly we produce triaxial haloes by fixing the ratio at $\pdm=0.88$, which matches the average found by \citet{chua2019} for Illustris Milky Way analogues. We also test an oblate halo model with $\pdm=1$ in Section~\ref{section:oblate}.

\subsection{Orbit library}\label{section:library}
We initialise the stellar orbit library by drawing phase space position samples from the density distribution defined by equation~\eqref{eq:rho_dpl} and Table~\ref{tab:halo_parameters}. The velocities of each particle are drawn from a Gaussian distribution, where the velocity dispersions are calculated from a spherical Jeans model with constant anisotropy $\betaan$. To match our observational anisotropy constraint, we set $\betaan=0.9$. Each orbit library uses a total of $N_\mathrm{orb}=5000$ initial conditions. Each orbit is then integrated from its initial phase space position in the model potential using \textsc{agama} \citep{agama}. The integration time for each particle is equal to $50\,T_\mathrm{circ}$, where $T_\mathrm{circ}$ is the approximate period of a circular orbit at the particle's energy. For an orbit initialised at $r\sim10$~kpc this corresponds to a total integration time of $\sim15$~Gyr, which is greater than the age of the GSE. This is therefore a sufficiently long integration time to capture variations on realistic timescales for the GSE. The orbit integrations are run in the $(X_*,Y_*,Z_*)$ reference frame aligned with the principal axes of the target stellar halo distribution.

\subsection{Orbit weight optimization}\label{section:optimization}
Following the generation of the orbit library, we optimize the weights $w_i$ of each orbit (indexed by $i$) to find the best match to the target density distribution (Section~\ref{section:data_density}) and anisotropy profile (Section~\ref{section:data_anisotropy}).

The density constraint uses the \textsc{DensitySphHarm} scheme implemented in \textsc{agama}. This quantifies the contribution $u_{i,n}$ of each orbit to a set of spherical harmonics (indexed by $n$). The radial dependence is represented by a set of 1st degree B-splines defined by a grid in $r$ \citep[these are $\wedge$-shaped functions spanning adjacent grid cells; e.g. see][]{vasiliev2020}. At present, \textsc{agama} only evaluates harmonics with even $l$ and non-negative even $m$. This corresponds to triaxial symmetry where the principal axes are aligned with the coordinate axes. Hence we are required to perform the fitting in the $(X_*,Y_*,Z_*)$ coordinate system aligned with the principal axes of the target distribution. The maximum order and degree of the harmonics are set to $(l_\mathrm{max},m_\mathrm{max})=(8,6)$. We use a non-uniform radial grid including the origin\footnote{Note that the `density' constraints are actually projections over finite volumes using the B-splines, so the diverging density at the origin is not a problem.} and 24 points in the range $6\leq r/\mathrm{kpc}\leq60$ \citepalias[the radial interval fitted by][]{han2022}. The target density distribution (Section~\ref{section:data_density}) is evaluated in the spherical harmonic basis to give a set of target components $U_n$. The aim of the optimization is to find a set of non-negative weights that most closely satisfies the constraints
\begin{equation}\label{eq:density_constraint}
    \sum_{i=1}^{N_\mathrm{orb}}w_i\,u_{i,n}=U_n
\end{equation}
for each $n$. For the above setup there is a total of $N_\mathrm{cons}=337$ of these constraints (allowing for the $r=0$ grid point having only a single harmonic with $l=m=0$).

The anisotropy profile constraint is provided via \textsc{KinemShell} in \textsc{agama}. This measures the density-weighted velocity dispersions $\rho\sigma_r^2$ and $\rho\sigma_t^2$, again projected onto a set of B-splines to represent the radial dependence \citep{agama}. Similarly to equation~\eqref{eq:density_constraint}, the projections of these quantities $U_n^r$ and $U_n^t$ on the $n$th radial basis element include $w_i$-weighted sums over all orbits. The constraints required to be satisfied are
\begin{equation}\label{eq:kin_constraint}
    0=2(1-\betaan)U_n^r - U_n^t,
\end{equation}
where $\betaan=0.9$ is the observation-based anisotropy. Unlike the density constraints in equation~\eqref{eq:density_constraint}, these are required to be satisfied exactly. This avoids the need to choose an arbitrary relative weighting of the two constraints, and due to the flexibility of the Schwarzschild model it should not compromise the density fit. We use 5 uniformly spaced radial grid points between 6 and 24 kpc, marked with triangles in Fig.~\ref{fig:beta_an_data}. These span the approximate range where the GSE anisotropy is well constrained (Section~\ref{section:data_anisotropy}).

In addition to satisfying density and kinematic constraints, Schwarzschild modelling typically employs regularization to prevent a small number of orbits dominating the fit. The optimization is performed by minimizing an objective function encoding the density constraint, with regularization provided by a term quadratic in the orbit weights. We use the objective function
\begin{equation}\label{eq:objective}
    \mathcal{Q}\equiv\frac{1}{M_*}\sum_{n=1}^{N_\mathrm{cons}}\left|\sum_{i=1}^{N_\mathrm{orb}}w_i\,u_{i,n}-U_n\right|+\frac{\lambda}{N_\mathrm{orb}}\sum_{i=1}^{N_\mathrm{orb}}\left[\frac{w_i}{M_*/N_\mathrm{orb}}\right]^2,
\end{equation}
where $\lambda=0.001$ is the regularization strength. We find that this is sufficiently large to give smooth solutions without affecting the quality of the fits. The denominators in equation~\eqref{eq:objective} ensure that each term is independent of the total mass $M_*$ and (roughly) the number of orbits $N_\mathrm{orb}$. The objective function is minimized subject to the kinematic constraint (equation~\ref{eq:kin_constraint}) and the total mass and non-negativity constraints,
\begin{align}
    \sum_{i=1}^{N_\mathrm{orb}}w_i&=M_*,\\
    w_i&\geq0\quad\forall\,i,
\end{align}
all of which must be satisfied exactly. The minimization is performed using the \textsc{solveOpt} function in \textsc{agama}, which uses the \textsc{CVXOPT} solver \citep{cvxopt}.

\begin{figure*}
  \centering
  \includegraphics[width=\textwidth]{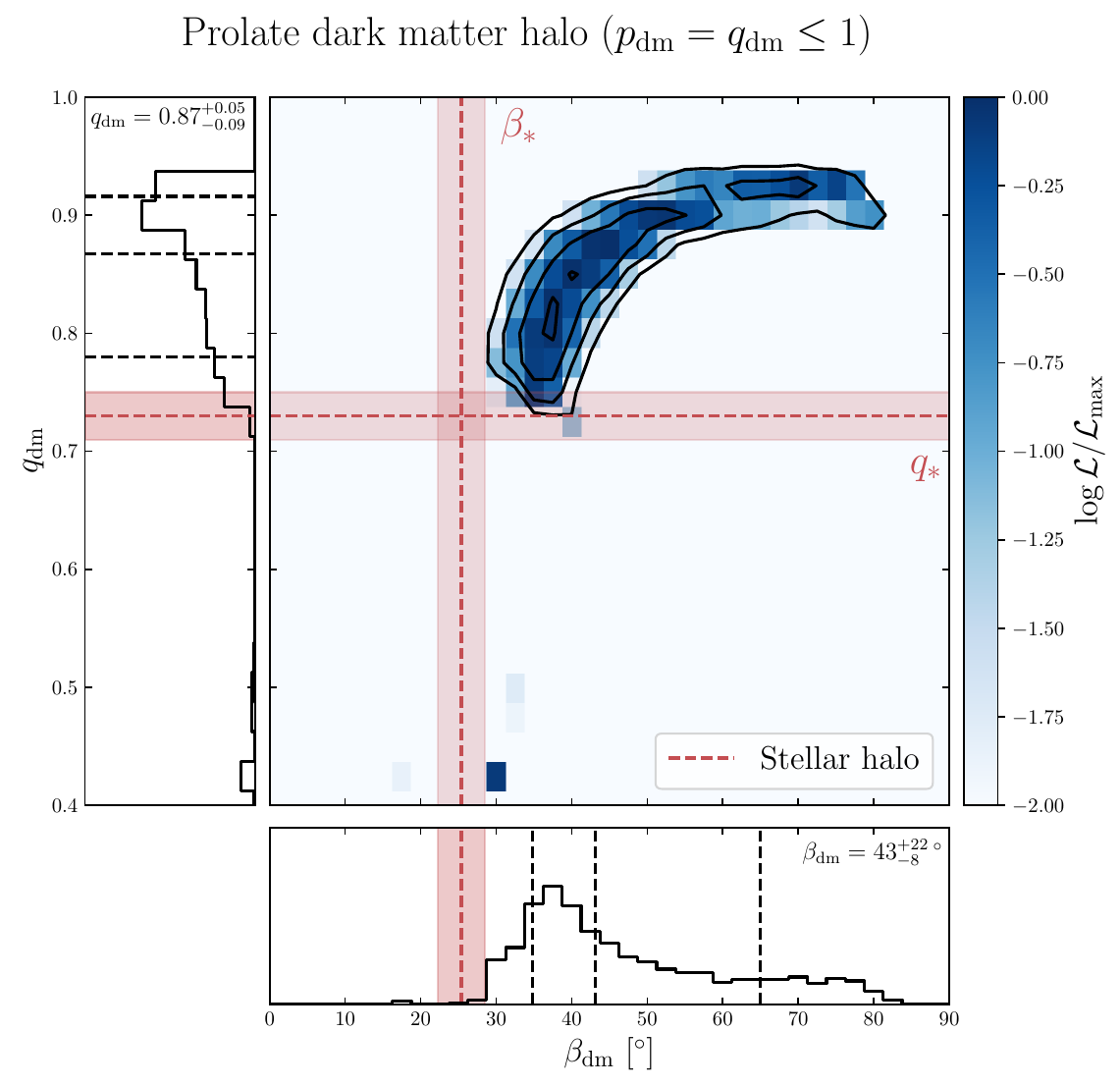}
  \caption{Results of the Schwarzschild fits in a prolate dark matter halo. The $x$ and $y$-axes correspond to the tilt $\betadm$ of the dark matter major axis with respect to the Galactic plane, and the short-to-long axis ratio $\qdm$ respectively. The blue pixels indicate the log-likelihood of the model. The black contours are calculated from a smoothed log-likelihood and placed at the $\mathrm{log}\,\mathcal{L}/\mathcal{L}_\mathrm{max}=-n^2/2$ levels, where $n\in\{0.5, 1.0, 1.5, 2.0\}$. The black histograms in the bottom and left-hand panels show the marginalized posteriors of $\betadm$ and $\qdm$, assuming uniform priors. Their medians and 16th/84th percentiles are marked with black dashed lines and printed in their respective panels. For comparison the red dashed lines and bands indicate the analogous quantities of the stellar halo fit by \citetalias{han2022}.}
   \label{fig:logL_grid_prolate}
\end{figure*}

\begin{figure*}
  \centering
  \includegraphics[width=\textwidth]{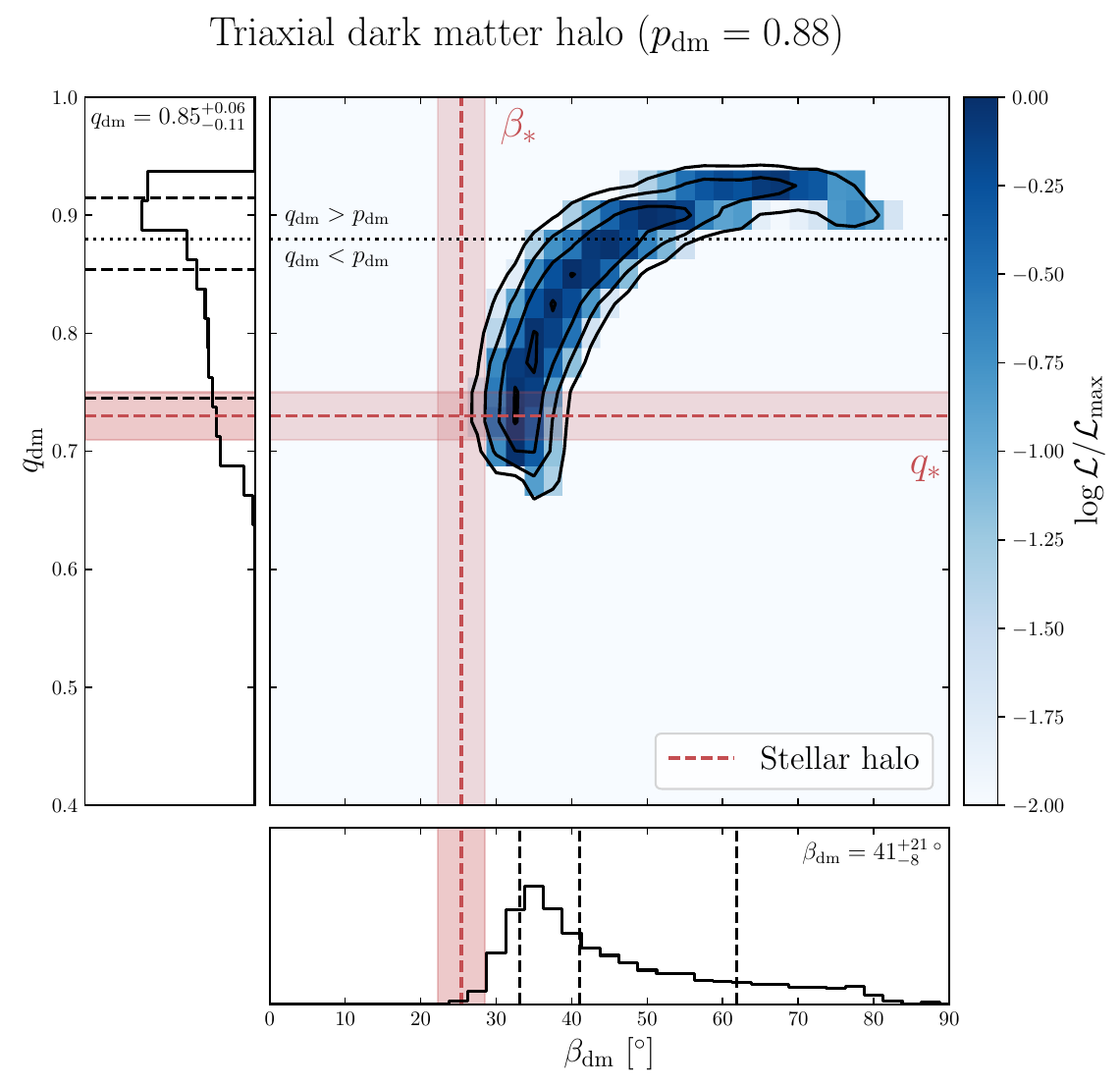}
  \caption{Like Fig.~\ref{fig:logL_grid_prolate}, but with a fixed axis ratio $\pdm=0.88$. The dotted line marks where $\qdm=\pdm$ (i.e. where the halo is prolate). The results are very similar to those for the prolate halo, except that slightly lower values of $\qdm$ are permitted ($\qdm\sim0.7$).}
   \label{fig:logL_grid_triaxial}
\end{figure*}

\subsection{Goodness of fit estimation}\label{section:goodnessoffit}
To place constraints on the potential using the Schwarzschild models we need a measure of goodness of fit. One option is the loss $\mathcal{Q}$ minimized in the fitting procedure to find the orbit weights $w_i$. However, this does not take into account uncertainties in the density fit or correlations between different harmonics. It is therefore unsuitable as a measure of how well the model matches observations. Furthermore we are only interested in finding potentials that are consistent with the overall shape and orientation of the triaxial stellar halo, not exactly matching its density at all positions. We therefore define a new goodness of fit measure that captures these global properties, to be used instead of $\mathcal{Q}$. This is the log-likelihood,
\begin{align}\label{eq:logL}
\begin{split}
    \mathrm{log}\,\mathcal{L}=-\frac{(\beta_\mathrm{data}-\beta_\mathrm{model})^2}{2\sigma_{\beta,\mathrm{data}}^2}&-\frac{(q_\mathrm{data}-q_\mathrm{model})^2}{2\sigma_{q,\mathrm{data}}^2} \\&-\frac{(p_\mathrm{data}-p_\mathrm{model})^2}{2\sigma_{p,\mathrm{data}}^2}.
\end{split}
\end{align}
Here $\beta$, $q$ and $p$ are measures of the tilt and axis ratios of the stellar halo, and the subscripts `data' and `model' indicate the observed density fit and Schwarzschild model respectively. However, the data shape parameters do \emph{not} correspond to those given in Table~\ref{tab:halo_parameters}. This is because those describe the shapes of isodensity surfaces, while in general we cannot uniquely calculate their equivalent values in a dynamical model (e.g. the tilt is likely to be radius-dependent). Instead we need a different set of parameters that can be compared like-for-like between the data and model. To do this we generate $N$-body snapshots of both the data and Schwarzschild model densities, with $10^6$ particles each. For each we calculate the inertia tensor
\begin{align}\label{eq:inertia}
    I_{ij}=\sum_{ij}^{6<r/\mathrm{kpc}<60}X_i X_j,
\end{align}
where $X_i$ are the Cartesian coordinates of the star particles in the $(X_*,Y_*,Z_*)$ frame, and the sum is over all particles between radii of 6 and 60~kpc. The axis ratios $q$ and $p$ are the ratios of the square roots of the eigenvalues, and $\beta$ is the angle between the $Z_*$-axis and the nearest eigenvector. The data uncertainties $\sigma_{\beta,\mathrm{data}}$, $\sigma_{q,\mathrm{data}}$, and $\sigma_{p,\mathrm{data}}$ are found by drawing 1000 samples of the data density from the uncertainties in $\beta_*$, $q_*$, and $p_*$ given in Table~\ref{tab:halo_parameters}, assuming that they are Gaussian distributed. In the posterior fit by \citetalias{han2022} these parameters are only weakly correlated, so we ignore correlations in our definition of the likelihood. We have checked that the variation in $q$ and $p$ between different $N$-body realisations is smaller (by a factor of $\gtrsim10$) than $\sigma_{q,\mathrm{data}}$ and $\sigma_{p,\mathrm{data}}$ at this resolution. The resultant means and uncertainties do not equal those in Table~\ref{tab:halo_parameters} because $I_{ij}$ is calculated in a spherical shell (in particular, the axis ratios are closer to unity).

We calculate the log-likelihood defined by equation~\eqref{eq:logL} for each model potential, and use this value as our goodness of fit measure. We note that solutions from Schwarzschild modelling are highly non-unique \citep{vasiliev2013}, so a given solution with high likelihood is not necessarily the only good fit to the data.

\begin{figure}
  \centering
  \includegraphics[width=\columnwidth]{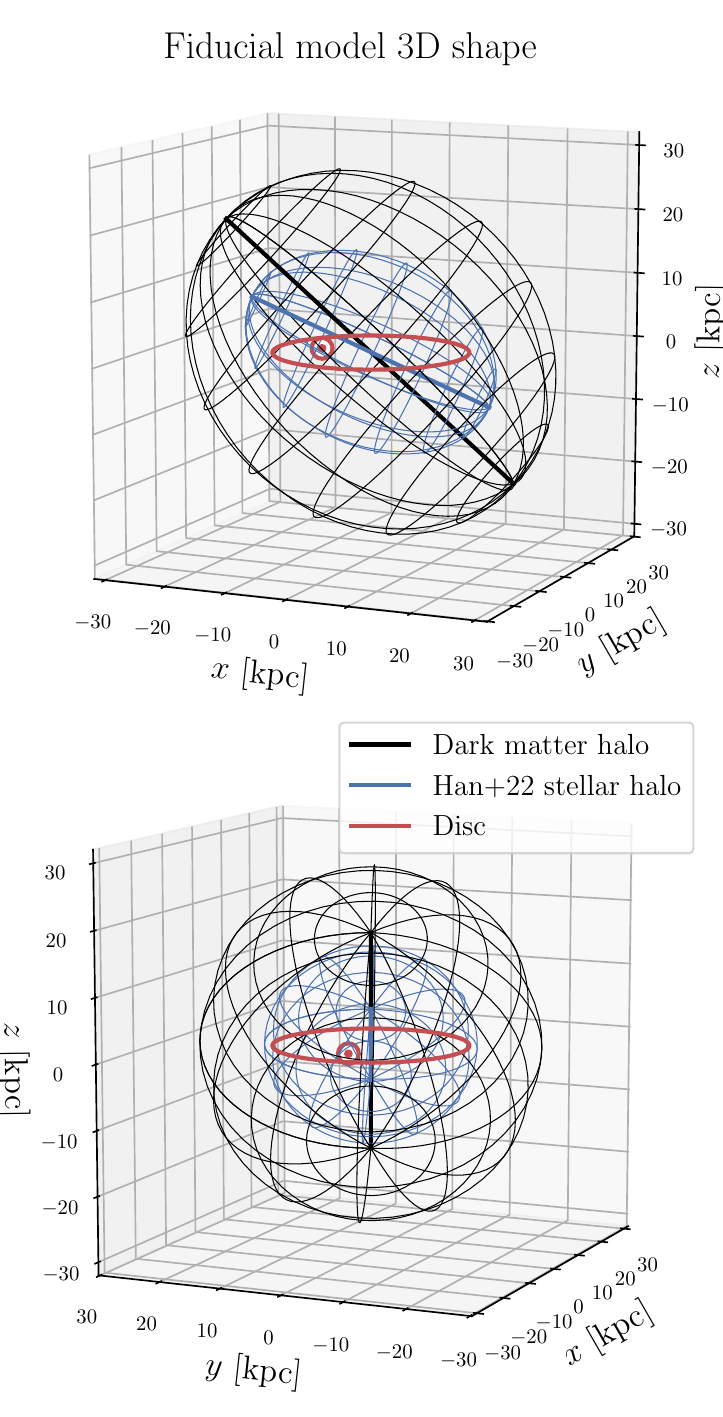}
  \caption{3D shape of the fiducial prolate dark matter halo compared to the \citetalias{han2022} stellar halo in the standard $(x,y,z)$ Galactocentric coordinate system. These are evaluated at ellipsoidal radii $r_\mathrm{dm}=30$~kpc and $r_*=20$~kpc respectively. Their major axes are shown as solid lines. The red $\odot$ symbol and circle mark the position of the Sun and the disc respectively.}
   \label{fig:3d_shape}
\end{figure}

\begin{figure}
  \centering
  \includegraphics[width=0.99\columnwidth]{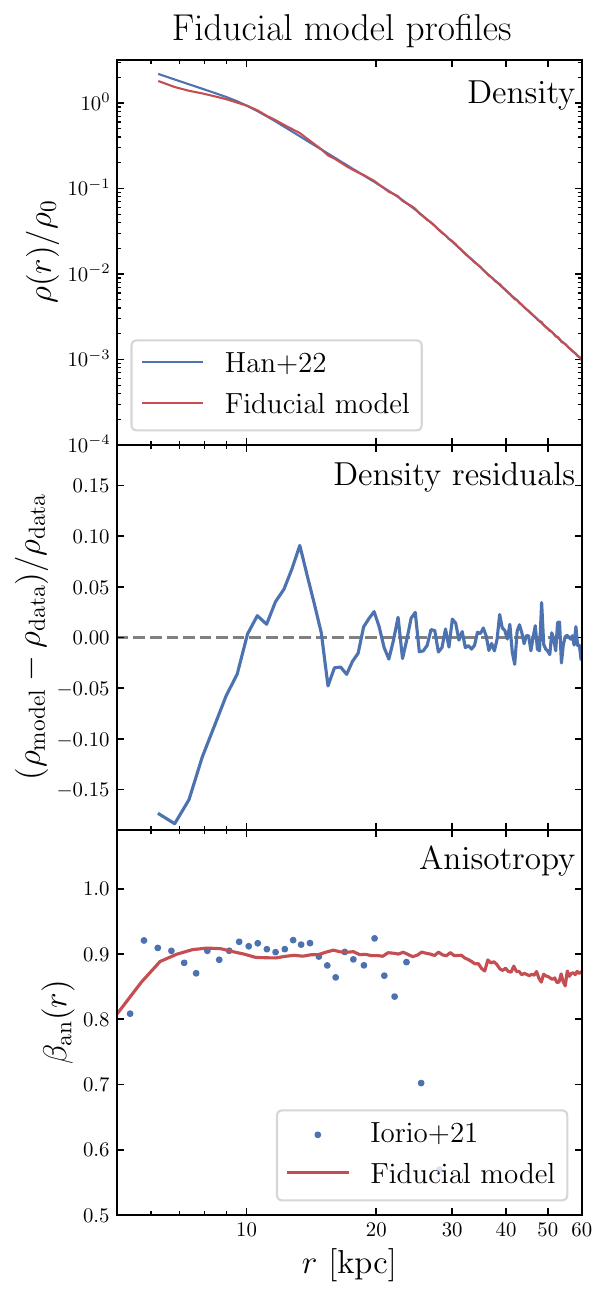}
  \caption{Radial profiles of the fiducial Schwarzschild model compared to the data. From top to bottom, the panels show the spherically averaged density profile $\rho(r)$, the fractional residuals of the model compared to the \citetalias{han2022} density fit, and the anisotropy profile $\betaan(r)$ compared to the \citet{iorio2021} RR Lyrae data. The model generally matches the data well; the fractional errors in density only exceed $10\%$ at radii of $r<8$~kpc.}
   \label{fig:rho_beta_profiles}
\end{figure}

\begin{figure*}
  \centering
  \includegraphics[width=\textwidth]{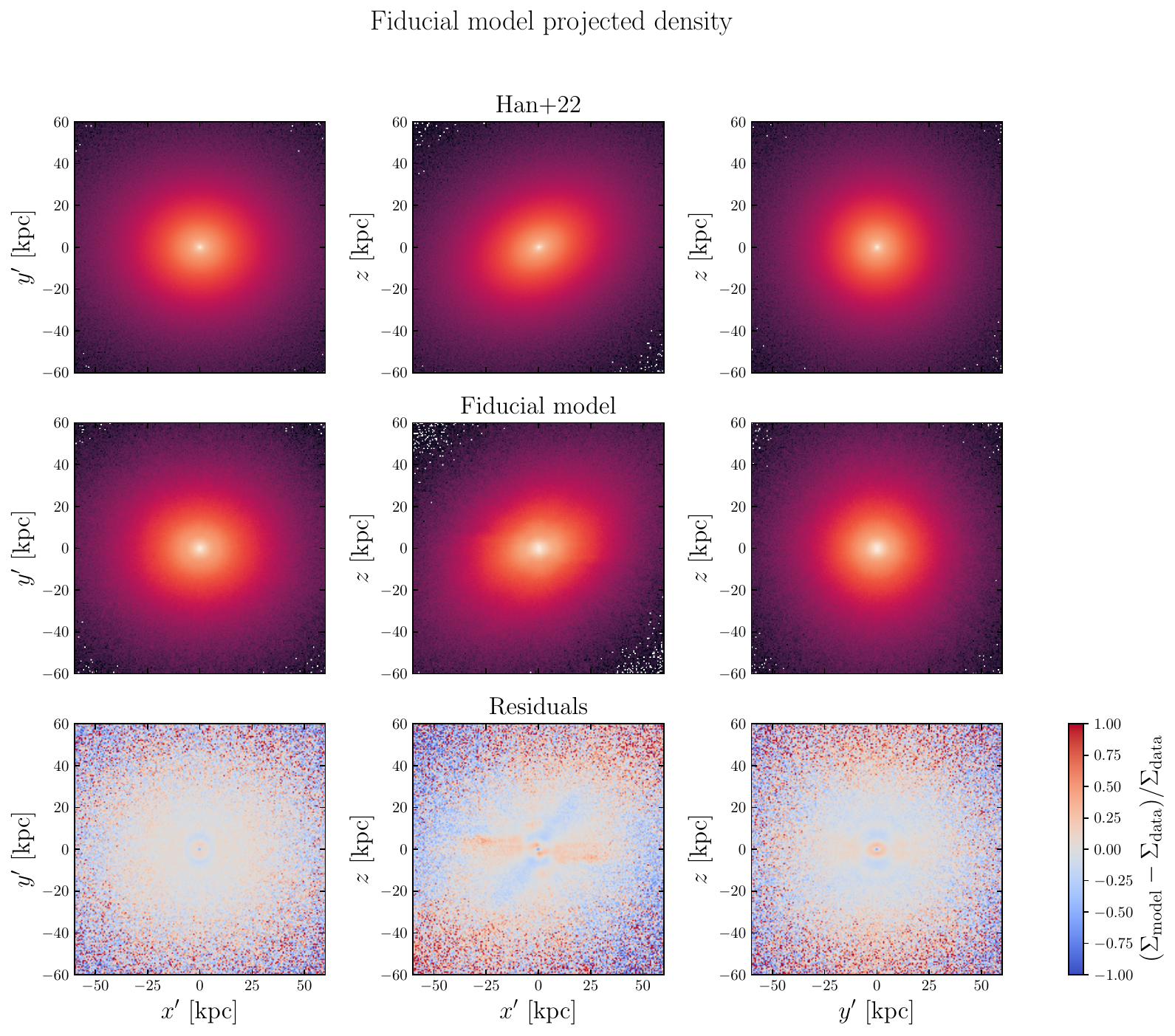}
  \caption{Projected density of the \citetalias{han2022} density (top row), our fiducial Schwarzschild model (middle row), and the fractional residuals (bottom row). From left to right, the columns show the $(x',y')$, $(x',z)$, and $(y',z)$ projections. The residuals are generally small, of order $\lesssim10\%$. The largest errors are seen in the $(x',z)$ projection (middle panel), which is the plane in which the halo is tilted relative to the disc.}
   \label{fig:density_residuals}
\end{figure*}

\begin{figure*}
  \centering
  \includegraphics[width=\textwidth]{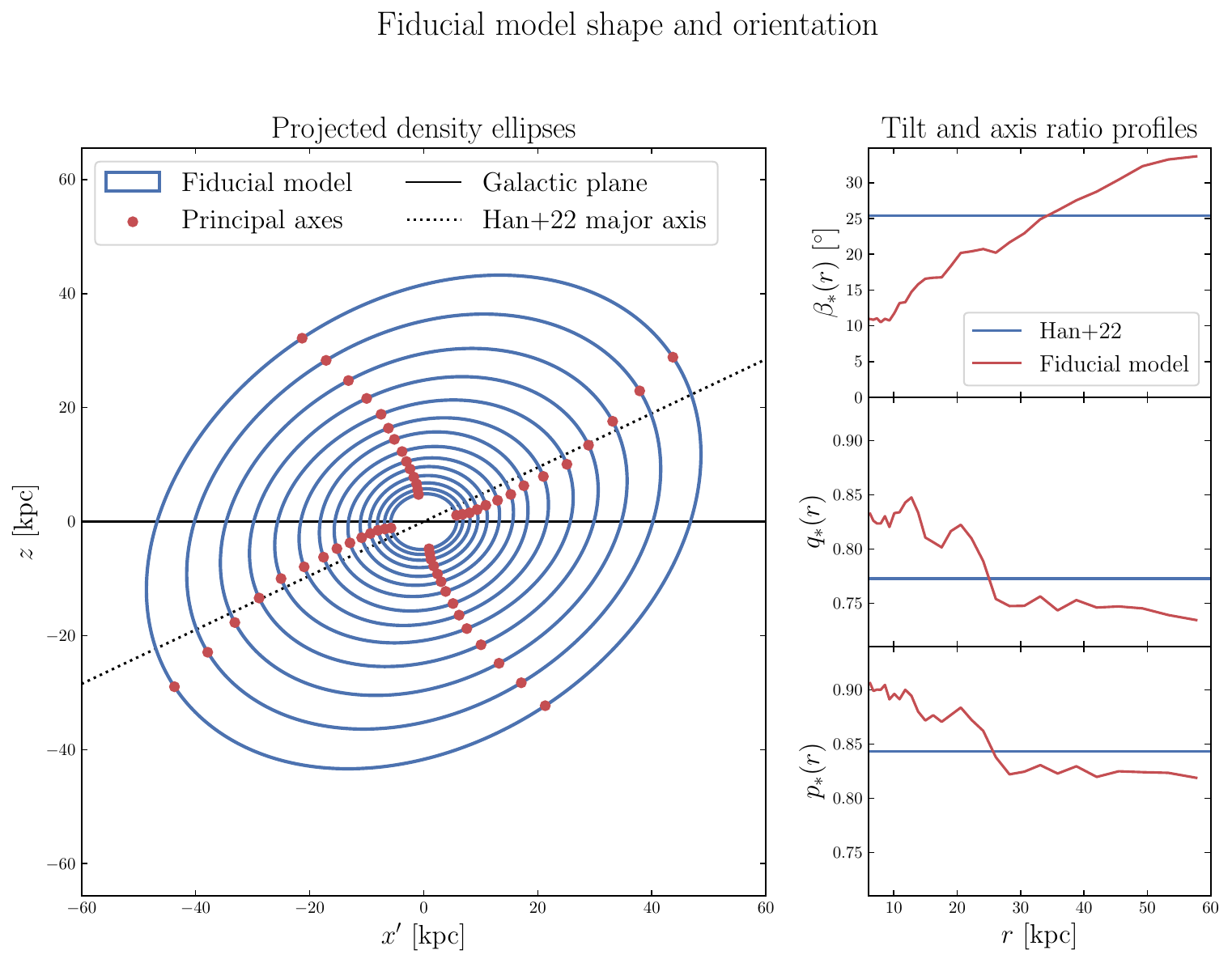}
  \caption{Shape and orientation of our fiducial model compared to the \citetalias{han2022} fit. \textbf{Left-hand panel:} ellipses illustrating the covariance of the fiducial model's projected density in the $(x',z)$ plane, in different radial bins. The circular points indicate the principal axes of each ellipse. The solid and dotted black lines mark the Galactic plane and the major axis of the \citetalias{han2022} density respectively. \textbf{Right-hand column:} radial dependence of the model's tilt (top panel) and axis ratios (middle and bottom panels). As $r$ increases, the model becomes more tilted and more elongated (i.e. less spherical).}
   \label{fig:axis_profiles}
\end{figure*}

\begin{figure*}
  \centering
  \includegraphics[width=\textwidth]{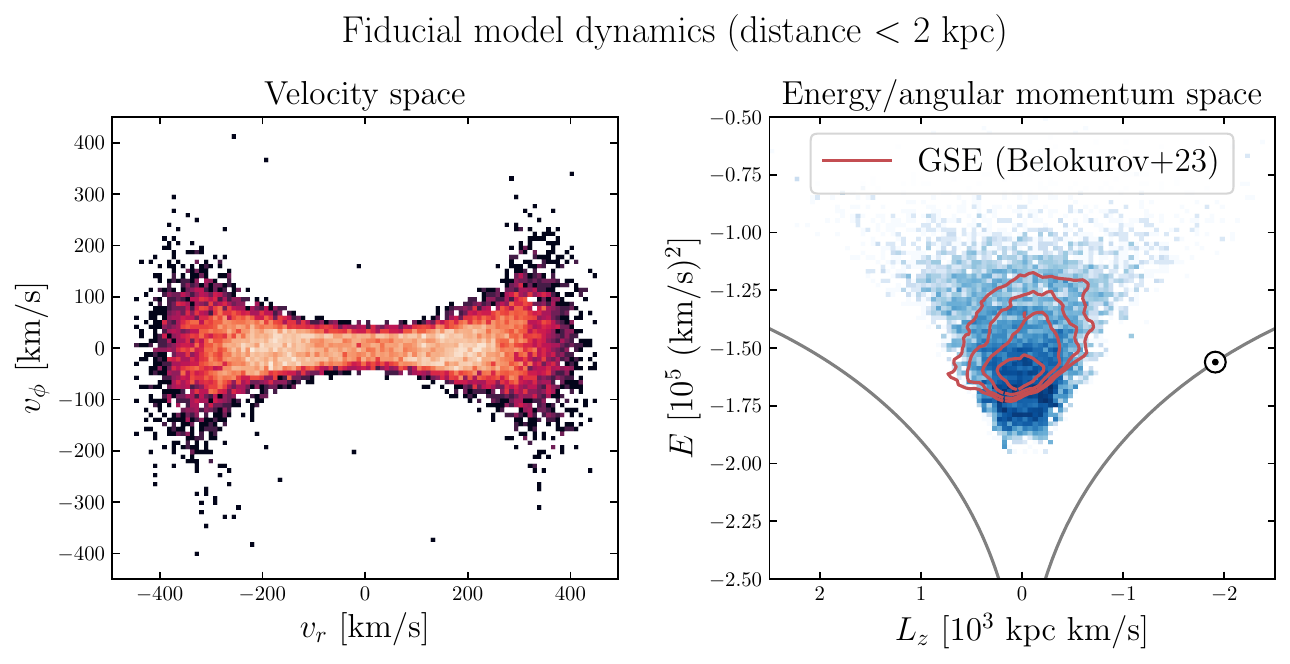}
  \caption{Dynamics of stars in the fiducial model less than 2~kpc from the Sun. \textbf{Left-hand panel:} velocity space in spherical coordinates, $(v_r,v_\phi)$. The stars are all on highly radial orbits, in a distribution closely resembling the GSE. \textbf{Right-hand panel:} energy vs angular momentum space $(L_z,E)$, with a circular orbit at the Sun's radius marked with a $\odot$ symbol. The histogram shows the fiducial model, while the contours indicate the density of the GSE in \textit{Gaia} DR3 derived by \citet{belokurov_chevrons}. There is a strong overlap between our model and the contours.}
   \label{fig:solar_neighbourhood}
\end{figure*}

\section{Results}\label{section:results}
\subsection{Dark matter halo constraints}
\begingroup
\begin{table}
    \centering
    \caption{Constraints on the tilt $\betadm$ and flattening $\qdm$ of the dark matter halo models. The two columns show results for prolate and triaxial (with $\pdm=0.88$) haloes. These are consistent with each other.}
    \renewcommand{\arraystretch}{1.5}
    \begin{tabular}{c|c|c}
    \hline
        Parameter & Prolate ($\pdm=\qdm$) & Triaxial ($\pdm=0.88$)\\
        \hline
        $\betadm$ & $43_{-8}^{+22}\,^\circ$ & $41_{-8}^{+21}\,^\circ$\\
        $\qdm$ & $0.87_{-0.09}^{+0.05}$ & $0.85_{-0.11}^{+0.06}$\\     
    \end{tabular}
    \label{tab:posteriors}
\end{table}
\endgroup
In Fig.~\ref{fig:logL_grid_prolate} we show the log-likelihood $\mathrm{log}\,\mathcal{L}$ of the prolate dark matter halo models over the grid of tilt angle $\betadm$ and axis ratio $\qdm$. The bottom and left-hand panels show the posterior distributions of $\betadm$ and $\qdm$ assuming uniform priors on both parameters. For comparison the red dashed lines indicate the tilt and flattening of the stellar halo from Table~\ref{tab:halo_parameters}, with the bands indicating their associated uncertainties. The black dashed lines mark estimates of the medians and 16th/84th percentiles of the posterior distributions. These are found by interpolating between the grid points, assuming that the probability distributions are flat around each point. These values are given in Table~\ref{tab:posteriors}. The tilt and axis ratio are $\betadm=43_{-8}^{+22}\,^\circ$ and $\qdm=0.87_{-0.09}^{+0.05}$ respectively. The dark matter halo is therefore expected to be both more tilted with respect to the disc and more spherical than the stellar halo (which has $\beta_*=(25.39\pm3.16)^\circ$, $q_*=0.73\pm0.02$). However, axis ratios $\qdm\geq0.95$ are ruled out by this model. This confirms the conclusion of \citet{han2022b} that a potential with a spherical dark matter halo cannot support a tilted triaxial stellar halo. This is as expected, since orbits in general axisymmetric potentials will precess around the symmetry axis at various rates \citep[e.g.][]{erkal2016}, thus erasing any elongated structure. There is a strong correlation between $\betadm$ and $\qdm$; as the halo becomes more spherical, the required tilt increases. This can be understood intuitively; a more spherical halo exerts a weaker torque on an orbit, so a greater tilt is required to balance the torque of the disc.

Fig.~\ref{fig:logL_grid_triaxial} shows the likelihood of the models with a triaxial dark matter halo, with fixed $\pdm=0.88$. These results are very similar to those for the prolate halo. The only significant difference is that slightly lower values of $\qdm$ are permitted, down to $\qdm\approx0.7$. This may be due to the intermediate axis ratio $\pdm$ being larger than that of the prolate model at the same value of $\qdm$ (for $\qdm<0.88$), so overall the halo is closer to spherical at a given $\qdm$. For this set of models the constraints on the dark matter halo are $\betadm=41_{-8}^{+21}\,^\circ$ and $\qdm=0.85_{-0.11}^{+0.06}$ (also given in Table~\ref{tab:posteriors}). The median model is therefore close to prolate, and the results of the two grids are fully consistent. In Section~\ref{section:discussion} we compare these results to a selection of recent studies and cosmological simulations. First we proceed to analyse the properties of the best-fitting stellar halo models.

\begin{figure}
  \centering
  \includegraphics[width=\columnwidth]{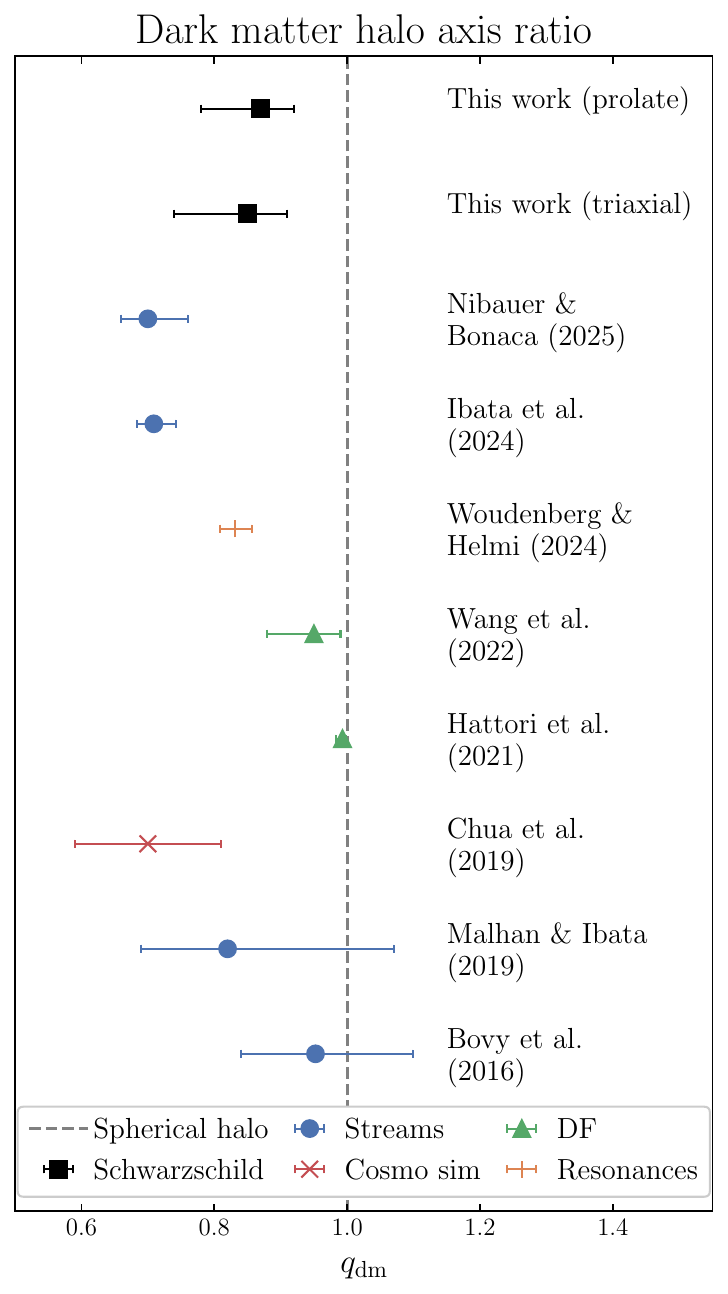}
  \caption{Constraints on the short-to-long axis ratio of the dark matter halo density distribution. The grey dashed line indicates a spherical halo. Our results are shown with black squares at the top. The other constraints come from stellar streams \citep{bovy2016,malhan2019,ibata2024,nibauer2025}, distribution function modelling of globlular clusters \citep{wang2022} and halo RR Lyrae stars \citep{hattori2021}, cosmological simulations of Milky Way analogues \citep{chua2019}, and resonant perturbations of substructure \citep{woudenberg2024}. Our results are consistent with each of these except the near-spherical constraint from \citet{hattori2021}.}
   \label{fig:q_dm_comparison}
\end{figure}

\begin{figure}
  \centering
  \includegraphics[width=\columnwidth]{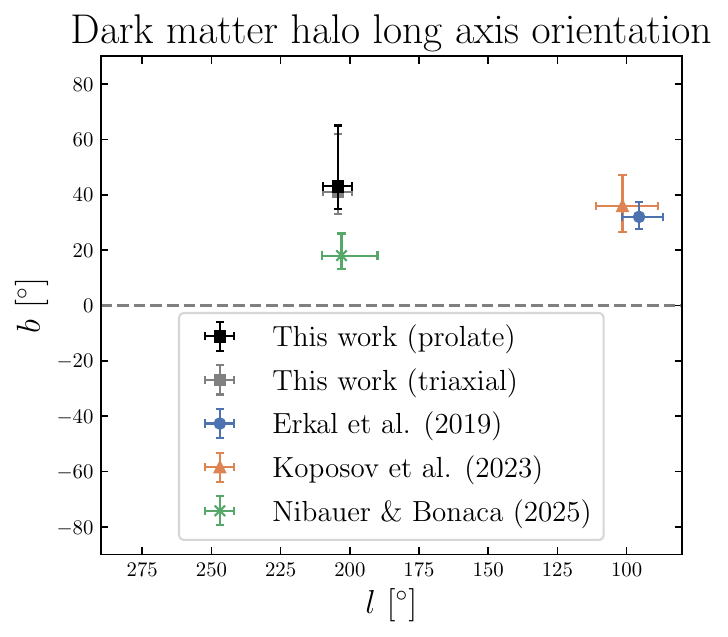}
  \caption{Constraints on the orientation of the dark matter halo's long axis in Galactic coordinates. For comparison we show three other estimates from models with (near-)prolate haloes. These are from the Orphan-Chenab stream \citep{erkal2019,koposov2023} and the GD-1 stream \citep{nibauer2025}.}
   \label{fig:orientation_comparison}
\end{figure}

\begin{figure}
  \centering
  \includegraphics[width=\columnwidth]{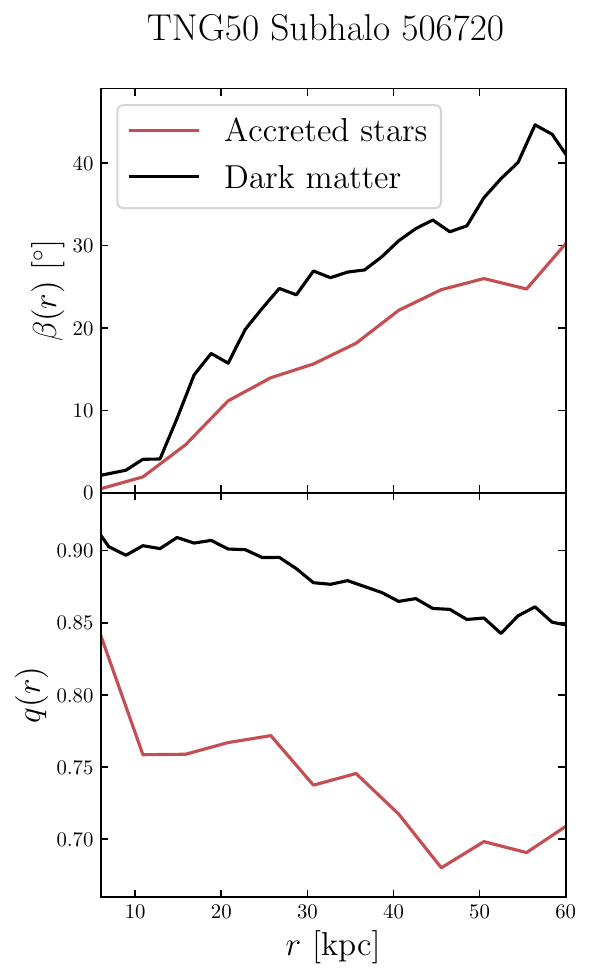}
  \caption{Tilt $\beta$ and short-to-long axis ratio $q$ profiles of the stellar and dark matter haloes in Subhalo 506720 of the TNG50 suite. In both cases the tilt increases with increasing radius. The dark matter halo is both more tilted and more spherical than the stellar halo at all radii.}
   \label{fig:tng50_profiles}
\end{figure}

\begin{figure}
  \centering
  \includegraphics[width=\columnwidth]{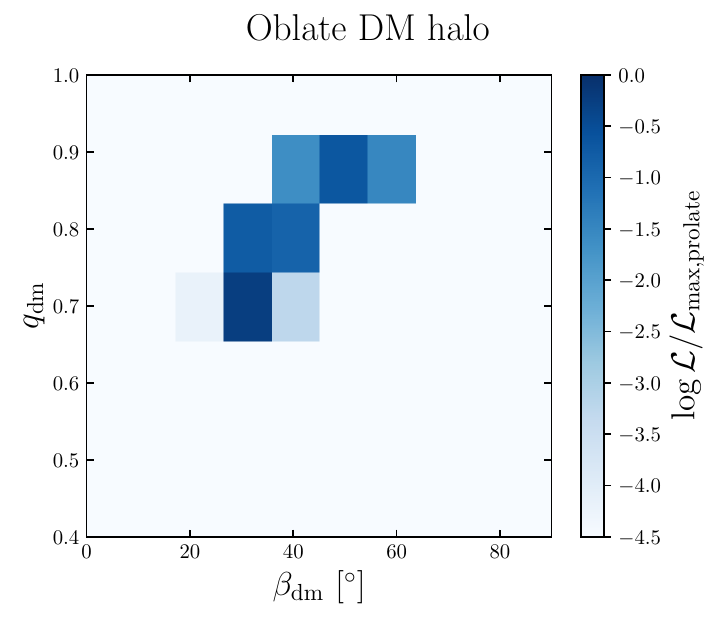}
  \caption{Like Fig.~\ref{fig:logL_grid_prolate}, but with an oblate halo ($\pdm=1$). The likelihood values are now normalised by the maximum likelihood of the prolate models shown in Fig.~\ref{fig:logL_grid_prolate}, so the colour scales are identical.}
   \label{fig:oblate_test}
\end{figure}

\begin{figure*}
  \centering
  \includegraphics[width=\textwidth]{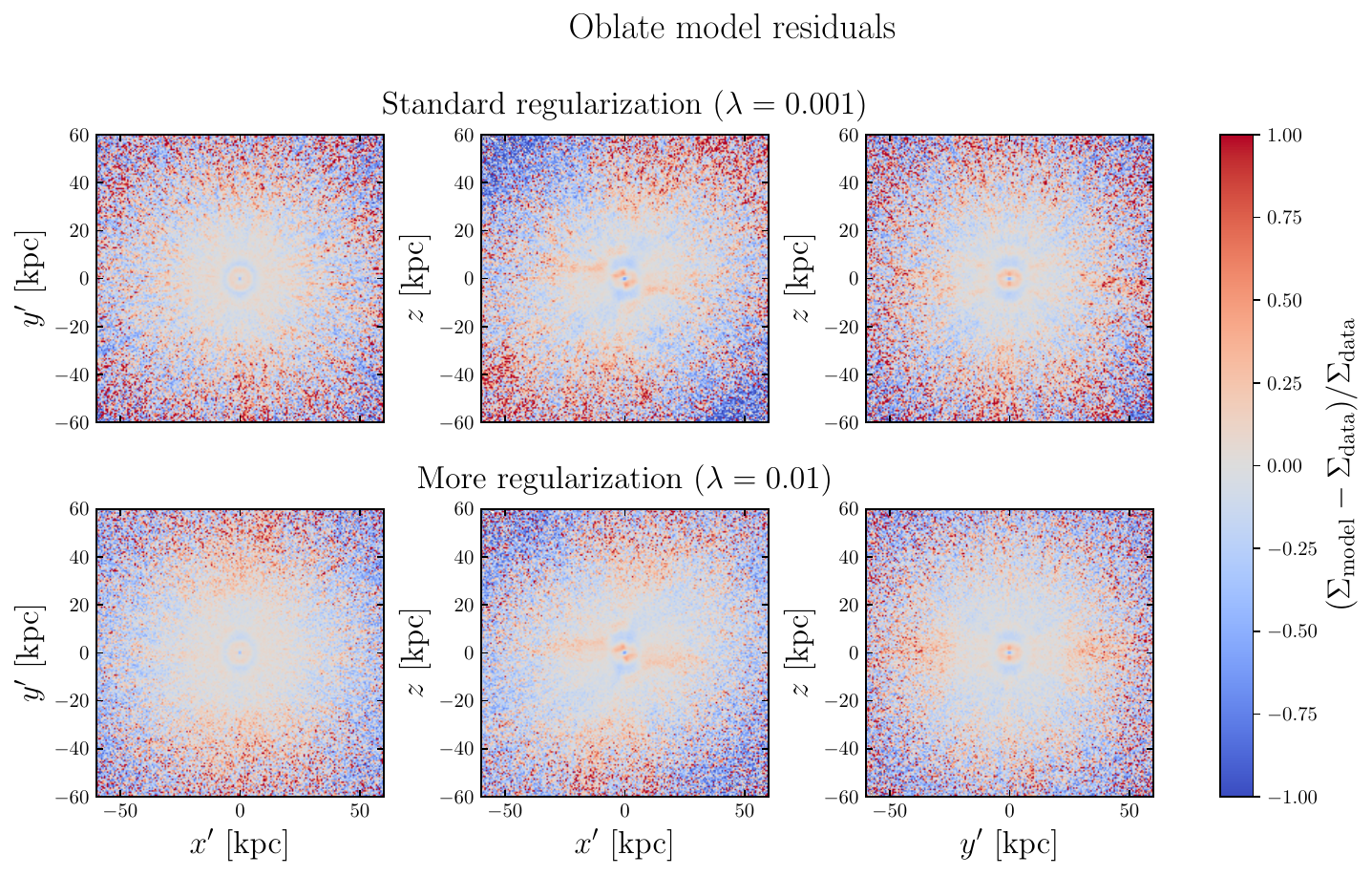}
  \caption{Surface density residuals of the oblate dark matter halo, with $\betadm=43.1^\circ$, $\qdm=0.867$, and $\pdm=1$. The top and bottom rows show fits with regularization parameters of $\lambda=0.001$ and 0.01 respectively. Compared to the prolate model shown in Fig.~\ref{fig:density_residuals}, the oblate model is either noisier (top row) or has much larger residuals at radii $r\gtrsim30$~kpc (bottom row).}
   \label{fig:oblate_residuals}
\end{figure*}

\begin{figure}
  \centering
  \includegraphics[width=\columnwidth]{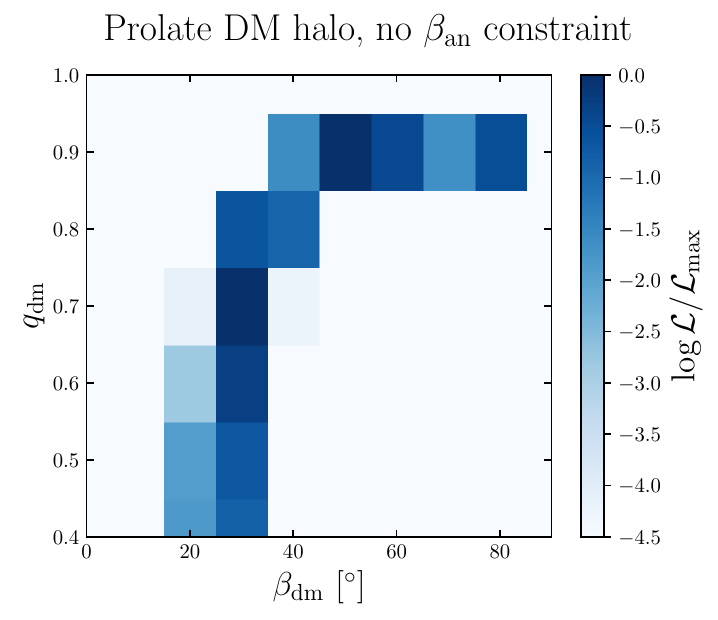}
  \caption{Like Fig.~\ref{fig:logL_grid_prolate}, but with Schwarzschild fits run with no constraint on the anisotropy $\betaan$. Much lower values of $\qdm$ are permitted when this constraint is removed.}
   \label{fig:anisotropy_test}
\end{figure}

\subsection{Best-fitting model properties}
As a fiducial model we select the median prolate dark matter model, with $\betadm=43.1^\circ$ and $\qdm=\pdm=0.867$. This is also not far from the median triaxial model with $\betadm=41^\circ$, $\qdm=0.85$, and $\pdm=0.88$. We show the 3D shape of the fiducial prolate dark matter halo compared to the \citetalias{han2022} stellar halo in Fig.~\ref{fig:3d_shape}. We re-run the Schwarzschild fit for this model at higher resolution, with $N_\mathrm{orb}=10000$ orbits integrated for time $100\,T_\mathrm{circ}$. We then generate an $N$-body snapshot of the fitted stellar halo with $10^7$ particles.

The radial density and anisotropy profiles of the fiducial model are compared to the data in Fig.~\ref{fig:rho_beta_profiles}. The top panel shows the spherically averaged density profile $\rho(r)$ of the model and the \citetalias{han2022} density. The two profiles generally match very well. The largest difference is at $r<10$~kpc, where the Schwarzschild model underestimates the density. This is shown more clearly in the middle panel, where we plot the fractional residuals of the model compared to the data. At most radii the error is less than $3\%$, and only exceeds $10\%$ at $r\lesssim8$~kpc. At these small radii the baryons dominate the Galactic potential rather than the tilted dark matter halo. It is therefore unsurprising that the Schwarzschild model deviates from the simple triaxial ellipsoidal structure of the target density distribution. This reflects the fact that the target has no radial dependence in its shape or orientation, so it is not necessarily consistent with dynamical equilibrium. We also note that the \citet{han2022} model is fitted to only data at high Galactic latitudes (i.e. excluding the disc), so it may not be sensitive to the shape near the Galactic plane. The bottom panel of Fig.~\ref{fig:rho_beta_profiles} shows the anisotropy profile $\betaan(r)$ of the fiducial model, with the points marking the RR Lyrae measurements from \citet{iorio2021}. The model matches these data points well, with the only exception being at $r\approx25$~kpc where \citet{iorio2021} measure a lower $\betaan$. However, the uncertainty of this measurement is large (see their Fig.~7), so our model is still consistent. The model also closely matches the measurements of \citet{lancaster2019} and \citet{bird2021} (see Fig.~\ref{fig:beta_an_data}). At $r>24$~kpc (outside the $\betaan$ constraints) the anisotropy remains approximately constant at $\betaan\approx0.9$. This is likely because the orbit library was generated from a spherical Jeans model with $\betaan=0.9$, and the model has sufficient freedom to fit the density profile without changing the anisotropy.

We show the projected density of the \citetalias{han2022} fit and our fiducial model in the first and second rows of Fig.~\ref{fig:density_residuals}. From left to right, the columns show projections in the $(x',y')$, $(x',z)$, and $(y',z)$ planes (i.e. those containing the principal axes of the stellar and dark matter haloes). The bottom row shows the fractional residuals between the data and model. In general the residuals are small, especially in the $(x',y')$ and $(y',z)$ panels. Broadly our Schwarzschild model thus successfully recovers the density profile of the \citetalias{han2022} fit. However, the match is not perfect. The residuals are larger in the $(x',z)$ projection (middle panel), the plane in which the halo is rotated relative to the disc normal. Compared to \citetalias{han2022} our model has a greater density in the disc plane ($z=0$), but a lower density around the major axis of the dark matter halo. This discrepancy extends up to a radius of $r\sim40$~kpc, and suggests that in the inner Galaxy our dynamical model is less tilted relative to the disc than the \citetalias{han2022} fit.

Despite the target density having fixed shape and orientation with radius, the above suggests that our dynamical model has some radial dependence. In Fig.~\ref{fig:axis_profiles} we show our fiducial model's shape and orientation as functions of radius. The large left-hand panel again shows the $(x',z)$ plane. We calculate the inertia tensors (see eq.~\ref{eq:inertia}) in this 2D space from particles in a series of logarithmically-spaced spherical shells. This is the `Local Shell Non-iterative Method' (LSNIM) described by \citet{emami2021}. We note that this returns more spherical shapes than alternative methods involving iterations. The ellipses are aligned with the matrix eigenvectors and have axis lengths $\sqrt{2\lambda_i}$, where $\lambda_i$ are the eigenvalues. We also mark the vertices (principal axis endpoints) of each ellipse with circular points. For comparison the disc plane and \citetalias{han2022} density major axis are shown with solid and dotted black lines respectively. This panel clearly shows that the tilt of the model stellar halo increases with radius, with a smaller tilt than the \citetalias{han2022} density at small radii. This follows the expectation from the residuals in Fig.~\ref{fig:density_residuals}. The ellipses also become visibly more elongated at larger radii, so the model stellar halo becomes more flattened with increasing radius. To quantify these effects we calculate the tilt $\beta_*(r)$ and axis ratios $\left(p_*(r),q_*(r)\right)$ from 3D coordinate covariance matrices in a series of spherical shells. These are shown as functions of $r$ in the right-hand column of Fig.~\ref{fig:axis_profiles}, with the \citetalias{han2022} model shown for comparison. The top panel confirms that the tilt of the Schwarzschild model halo increases steadily with radius, matching the \citetalias{han2022} value at an intermediate radius of $r\approx35$~kpc. The axis ratios both decrease with radius, crossing the \citetalias{han2022} values at $r\approx25$~kpc. This radius may be related to the extent to which the disc's potential is dominant. At smaller (larger) radii our model is more spherical (more elongated) than the target density. Hence while the dynamical model is a good fit to the \citetalias{han2022} density in an averaged sense, both its shape and orientation are functions of radius. This suggests that future efforts to fit the density of the tilted stellar halo should allow more flexibility in their models' radial dependence. However, we also note that since \textsc{agama} only supports even spherical harmonics in its Schwarzschild machinery, the axis ratio of the model may differ from that of the target without being penalised. This may be responsible for some of the discrepancy.

We show the Solar neighbourhood kinematics and dynamics of the fiducial model in Fig.~\ref{fig:solar_neighbourhood}. We select particles with Heliocentric distances less than 2~kpc, where the Sun is located at Galactocentric coordinates $(-8.2,0,0)$~kpc. The left-hand panel shows the velocity space in spherical coordinates $(v_r,v_\phi)$. As expected, the distribution of stars closely resembles the GSE \citep{belokurov2018}, with almost all stars on highly radial orbits. The distribution has peaks at $v_r=\pm200$~km/s rather than $v_r=0$. The observed GSE similarly shows residual excesses at large positive and negative $v_r$ when fitted with a single Gaussian \citep{belokurov2018}. The 2D histogram in the right-hand panel shows the energy and angular momentum $(L_z,E)$. The contours indicate the distribution of the GSE derived from \textit{Gaia} Data Release 3 \citep[DR3;][]{gaia_dr3} by \citet{belokurov_chevrons}. Again there is a close correspondence between the dynamical model and the observed GSE, though our model extends to slightly lower energy. This may be related to the fact that the \citetalias{han2022} density is only fitted to stars at $r>6$~kpc, so our model might overestimate the number of stars at small radii and low energies. Note that the selection functions of the two distributions are not identical, so they are not expected to exactly match. The narrow width of the distribution in $L_z$ is due to the high anisotropy constraint. The \citet{belokurov_chevrons} distribution also shows some asymmetry, with net positive (negative) $L_z$ at higher (lower) energy. This may be due to resonant or chaotic perturbations from the Galactic bar \citep{dillamore2023,dillamore2024,dillamore2025b,woudenberg2025}, and suggests that these effects should be considered in future iterations of our model. 

From Figs.~\ref{fig:rho_beta_profiles}-\ref{fig:solar_neighbourhood} we conclude that our fiducial dynamical model is a good fit to observations of the GSE's density and kinematics.

\section{Discussion}\label{section:discussion}
\subsection{Comparison with previous studies}
We show our estimates of the dark matter halo's short-to-long axis ratio compared to a selection of recent constraints in Fig.~\ref{fig:q_dm_comparison}. These come from stellar stream modelling \citep{bovy2016,malhan2019,ibata2024,nibauer2025}, distribution function modelling of globular clusters \citep{wang2022} and RR Lyrae stars \citep{hattori2021}, Milky Way analogues in cosmological simulations \citep{chua2019}, and resonances \citep{woudenberg2024}. Our results are consistent with each of these previous constraints except \citet{hattori2021}, since we rule out near-spherical haloes. We note that the differences between the various constraints may be partly due to their use of tracers at different Galactocentric radii. As Fig.~\ref{fig:axis_profiles} suggests, the axis ratios and orientations of a dynamically consistent halo (stellar or dark matter) are likely to be highly radius-dependent, so it is unsurprising that there is some variation between the different studies.

Our constraints on the orientation of the dark matter halo's long axis are shown in Galactic coordinates $(l,b)$ in Fig.~\ref{fig:orientation_comparison}. The error bars in $l$ have length equal to the uncertainty of the stellar halo yaw angle $\alpha_*$ (Table~\ref{tab:halo_parameters}). For comparison we show the prolate models used to fit the Orphan-Chenab (OC) stream by \citet{erkal2019} and \citet{koposov2023}, and the near-prolate solution inferred from accelerations along the GD-1 stream by \citet{nibauer2025} (their `Mode 1'). We do not include models with (near-)oblate haloes such as that of \citet{vasiliev_tango}. Compared to the axis ratio constraints in Fig.~\ref{fig:q_dm_comparison}, there is poor consistency among the orientation estimates. While the OC stream results favour a long axis pointing into the northern Galactic hemisphere at $l\approx100^\circ$, \citet{nibauer2025} prefer it to be at $l\approx200^\circ$. This is consistent with the orientation of the stellar halo as constrained by \citetalias{han2022}, and therefore with our own estimates. Note however that we favour a greater tilt angle than \citet{nibauer2025}, as well as an axis ratio closer to unity (see Fig.~\ref{fig:q_dm_comparison}). Some of the discrepancies between the estimates may be related to time dependence. While our method constrains the orientation required for long-term equilibrium, the OC stream is highly susceptible to perturbations from the LMC over timescales of $\sim1$~Gyr \citep{erkal2019,koposov2023}. In fact the long axes of the prolate haloes from \citet{erkal2019} and \citet{koposov2023} do point approximately in the direction of the LMC. This lead \citet{erkal2019} to suggest that their MW models are compensating for some incorrect modelling of the perturbation, such as not considering deformations of the halo by the LMC \citep{lilleengen2023,koposov2023}. GD-1 is much less perturbed by the LMC, so the constraint by \citet{nibauer2025} is less likely to be influenced by this effect. In summary, while our results combined with others provide evidence for a mildly aspherical and tilted dark matter halo, its orientation remains very poorly constrained.

\subsection{Comparison with cosmological simulations}
Our estimates for the tilt and axis ratio of the Milky Way's dark matter halo can be compared with predictions from cosmological simulations run with a $\Lambda$CDM cosmological model. We have found that the tilt of the halo is $\betadm=43_{-8}^{+22}\,^\circ$. This is in excellent agreement with \citet{tenneti2016}, who found that disc galaxies in the Illustris suite \citep{illustris} with stellar masses in the range $10^{10.5-12}\mathrm{M}_\odot$ have a mean dark matter halo tilt of $46.5^\circ$. Our estimate is slightly larger than expected from the TNG50 suite \citep{tng50,nelson2019,pillepich2019}, as found by \citet{han2023}. They studied the sample of Milky Way analogues from \citet{pillepich2024}, defined as having stellar masses in the range $10^{10.5-11.2}\mathrm{M}_\odot$, a disc-like morphology, and a Milky Way-like environment on Mpc scales. They found that only $15\%$ of Milky Way analogues have dark halo tilts exceeding $40^\circ$. It is also larger than expected compared to the stellar halo tilt. While we estimate that the difference in tilt between the stellar and dark matter haloes is $\betadm-\beta_*\approx18_{-9}^{+22}\,^\circ$, \citet{han2023} found that the two tilts are on average approximately equal (albeit with considerable scatter). However, we note that these were calculated from inertia tensors in a fixed radial range, whereas in reality the tilt is likely to be radius-dependent. Hence the results may not be straightforward to compare. They also did not condition on merger history or baryonic mass fraction, which are likely to affect the dark matter halo structure \citep[e.g.][]{drakos2019}. More studies of tilted stellar haloes in cosmological simulations are needed to assess whether our results are consistent with the closest Milky Way analogues.

We have estimated that the short-to-long axis ratio of the dark matter halo is $\qdm=0.87_{-0.09}^{+0.05}$. \citet{chua2019} measured the shapes of dark matter haloes in the Illustris simulations \citep{illustris}. In Milky Way-mass haloes the two axis ratios were found to be $\qdm=0.70\pm0.11$ (which we show in Fig.~\ref{fig:q_dm_comparison}) and $\pdm=0.88\pm0.10$. Our estimated flattening is therefore just about consistent with the range found by \citet{chua2019}, although we favour a somewhat more spherical halo. \citet{emami2021} measured the dark matter halos shapes of Milky Way analogues in the TNG50 suite using three different algorithms. While these gave somewhat different measurements of the flattening, their estimates span the approximate range $\qdm\approx0.7-0.9$, in agreement with our estimate for the Milky Way.

We also compare our results with an individual Milky Way analogue from the TNG50 suite. \citet{wu2022} studied the effect of the GSE on the shape of the stellar halo, and identified two close analogues of the Milky Way suitable for this purpose. We inspect both of these and find that one, Subhalo 506720, has a tilted stellar halo. This experienced a major merger $\sim8.4$ Gyr ago, similar to the GSE (though we note that its stellar mass is somewhat higher at $1\times10^{10}\mathrm{M}_\odot$). From this galaxy we select a stellar halo population comprised of all stars flagged as accreted (i.e. born outside the host galaxy). As in our fiducial model in Fig.~\ref{fig:axis_profiles}, we calculate the orientations and axis ratios of the accreted stellar and dark matter haloes in a series of radial bins. The tilt $\beta$ is defined as the angle between the disc's angular momentum axis and the halo's short axis, while the flattening $q$ is the ratio of the short and long axes. These are shown in Fig.~\ref{fig:tng50_profiles}. The top panel shows that the tilt of the dark matter halo exceeds that of the stellar halo at all radii, as expected from our results (Figs.~\ref{fig:logL_grid_prolate} and \ref{fig:logL_grid_triaxial}). Similarly the dark matter halo is less flattened than the stellar halo at all radii (bottom panel), again in line with our estimates for the Milky Way. With increasing radius both the stellar and dark matter haloes become more tilted and more flattened. This is in line with the predictions of our dynamical model for the stellar halo (Fig.~\ref{fig:axis_profiles}). The radial dependence of the dark matter halo suggests that future developments of our work should allow more flexible potentials models, such as by fitting a Schwarzschild model to the dark matter halo itself \citep[e.g. analogous to][]{piffl2015,binney2015}. Overall Fig.~\ref{fig:tng50_profiles} suggests that our results are reasonable for Milky Way-like galaxies in $\Lambda$CDM cosmological simulations with tilted haloes. A future investigation using a large sample of Milky Way analogues will allow our results to be compared with $\Lambda$CDM predictions more thoroughly.

\subsection{Oblate dark matter haloes}\label{section:oblate}
In Section~\ref{section:schwarzschild} we chose to only consider triaxial or prolate dark matter haloes. We now attempt to run fits to the \citetalias{han2022} stellar halo in an oblate dark matter halo, with $\pdm=1$. We use a coarser grid in $\betadm$ and $\qdm$, with a quadrupled grid step in each dimension. We show the results of these fits over the grid in Fig.~\ref{fig:oblate_test}. To allow comparison with the prolate models, we normalise the likelihoods by the maximum likelihood of the prolate models in Fig.~\ref{fig:logL_grid_prolate}. The colour scales of Figs.~\ref{fig:logL_grid_prolate} and \ref{fig:oblate_test} are therefore identical. At first glance the oblate models appear to perform almost as well as the prolate models. They provide similar constraints on $\betadm$ and $\qdm$, with only a slightly lower maximum likelihood. We show the surface density residuals of the estimated median model with $\betadm\approx35.9^\circ$ and $\qdm\approx0.806$ in the top row of Fig.~\ref{fig:oblate_residuals}\footnote{This has a higher likelihood than the maximum likelihood model in the grid at $\betadm=30^\circ$, $\qdm=0.7$.}. The residuals are visibly larger and noisier than in the median prolate model in Fig.~\ref{fig:density_residuals}, especially at larger radii ($r\gtrsim30$~kpc). We have found that this is also true in the radial density profile, with noisy residuals of $\gtrsim5\%$ at a wide range of radii. Since these are noisy fluctuations rather than global errors they are not reflected in the likelihood. They indicate that a large proportion of the mass is concentrated in a small number of orbits, despite the regularization intended to ensure a smooth solution (see Section~\ref{section:schwarzschild}). To test whether this problem can be solved by increasing the regularization, we also run a fit with the same parameters but with $\lambda=0.01$ (i.e. increased by a factor of 10). The residuals of this model are shown in the bottom panel of Fig.~\ref{fig:oblate_residuals}. While it is less noisy, this model is visibly a poorer fit to the global shape of the potential. This is most obvious in the left-hand panel, where the density of the model is larger than the target distribution along the $y'$-axis (i.e. the intermediate axis of the ellipsoid). These properties may be understood in terms of the potential's possible orbits. While potentials with prolate and triaxial haloes can support box orbits elongated along their major axes, these are not likely to exist if the halo is oblate \citep{binney_tremaine}. Hence it struggles to support a set of orbits that remain elongated along the major axis of the \citetalias{han2022} halo. Indeed we find that the stellar halo's intermediate axis ratio $p$ is typically larger than that of the fiducial model in the prolate halo. The solution with $\lambda=0.001$ also has a higher anisotropy ($\betaan\approx0.93$) at large radii than the fiducial model. This suggests that the triaxial structure can only be maintained with a significant fraction of extreme radial orbits. The resultant anisotropy of $\betaan>0.9$ significantly exceeds estimates for the stellar halo at large radii \citep[e.g.][]{bird2021,han2024,chandra2025}. In summary we find that fits in oblate dark matter haloes are either noisy (with low regularization) or not sufficiently elongated (with higher regularization). We conclude that while an oblate dark matter halo cannot be ruled out, it is less likely to be able to support a triaxial stellar halo in equilibrium.

\subsection{Effect of the anisotropy constraint}
In our fits in Section~\ref{section:schwarzschild} we enforced that the models have anisotropy $\betaan=0.9$ between $r=6$ and 24~kpc. This ensures that the model kinematics are consistent with measurements of the GSE. In this section we investigate how the results are affected when this constraint is relaxed. We re-run the prolate dark matter halo fits over the coarse grid used in Section~\ref{section:oblate}. We follow an identical procedure to Section~\ref{section:schwarzschild} except that the constraint defined by equation~\eqref{eq:kin_constraint} is removed. The log-likelihood is shown over the grid in Fig.~\ref{fig:anisotropy_test}. While the high likelihood regions are still restricted to a narrow band in this plane, they extend to much lower values of $\qdm$ than in Fig.~\ref{fig:logL_grid_prolate}. The anisotropy constraint was therefore responsible for ruling out the most elongated (least spherical) dark matter halo models. To determine why, we have examined the contributions to the likelihood from the three parameters $\{\beta_\mathrm{data},q_\mathrm{data},p_\mathrm{data}\}$ when the anisotropy constraint is used. At $\qdm\lesssim0.7$ the likelihood is heavily penalised by the shape parameters, $p_\mathrm{data}$ and especially $q_\mathrm{data}$. In these cases the dynamical model is more elongated and flattened than the target density. The observed shape and high anisotropy of the GSE therefore appear to place a limit on how elongated the dark matter halo can be.

\section{Conclusions}\label{section:conclusions}
We have fitted Schwarzschild (orbit-superposition) models to the \textit{Gaia} Sausage-Enceladus (GSE) component of the Milky Way's stellar halo. This allows us to constrain the orientation and shape of the dark matter halo, assuming that a) the GSE is in equilibrium, and b) the dark matter halo's major axis is coplanar with that of the stellar halo and the disc normal. Our findings are summarised below.

\begin{enumerate}[label=\textbf{(\roman*)}]
    \item The observed shape and anisotropy of the GSE are consistent with equilibrium in a tilted non-spherical dark matter halo. It is therefore possible that the GSE debris has remained close to its present-day configuration for the $\gtrsim8$~Gyr since its merger. This holds promise that it can be used to constrain the geometry of the merger \citep{naidu2021} with the help of cosmological simulations for comparison \citep[e.g.][]{fattahi2019,dillamore2022}.

    \item If the halo is prolate we estimate its long axis to be inclined at an angle of $\betadm=43_{-8}^{+22}\,^\circ$ to the Galactic plane, which is $\approx18^\circ$ greater than the stellar halo. The short-to-long axis ratio of its density distribution is $\qdm=0.87_{-0.09}^{+0.05}$. This is somewhat more tilted and less flattened than is typically found in $\Lambda$CDM cosmological simulations, though still consistent. Triaxial dark matter halo models with one fixed axis ratio give similar results. Our value for the short-to-long axis ratio is in excellent agreement with previous constraints. However, an oblate halo with the same tilt and flattening does not permit a good dynamical model of the GSE. Oblate dark matter haloes therefore appear to be inconsistent with a near-prolate tilted GSE in equilibrium. Spherical haloes are ruled out by our models, confirming the findings of \citet{han2022b}.

    \item More elongated dark matter haloes (i.e. with smaller $\qdm$) are ruled out by the constraint that the GSE has a high anisotropy of $\betadm\approx0.9$. Much lower values of $\qdm$ are permitted when this constraint is removed, which also allows the long axis to be inclined at a smaller angle to the disc. This demonstrates the importance of including kinematic constraints when modelling the stellar halo.
    
    \item We produce an $N$-body snapshot of the stellar halo Schwarzschild model in the median prolate dark matter halo, with $\betadm=43.1^\circ$ and $\qdm=\pdm=0.867$. This is generally a good fit to observations of the GSE, including the \citetalias{han2022} density and anisotropy measurements \citep{lancaster2019,iorio2021,bird2021}. However, unlike the target \citetalias{han2022} density distribution the shape and orientation are radius-dependent. The tilt increases with radius from $\betadm\approx10^\circ$ at $r=6$~kpc to $\approx35^\circ$ at $60$~kpc. The halo also becomes spherical with increasing radius, with the short-to-long (intermediate-to-long) axis ratio decreasing from $q\approx0.85$ to $\approx0.75$ ($p\approx0.9$ to $\approx0.82$) across the same radial range. A close Milky Way analogue from the TNG50 cosmological simulation suite shows the same behaviour of the stellar halo. Its dark matter halo is more inclined and more spherical than the stellar halo, which is also consistent with our results.

    \item The kinematics $(v_r,v_\phi)$ and integrals of motion $(L_z,E)$ of this model closely resemble those of the GSE in the Solar neighbourhood. The model is therefore an excellent equilibrium dynamical representation of the GSE that takes into account its observed tilted triaxial structure. This represents an upgrade compared to the standard axisymmetric models of the stellar halo defined by distribution functions. It provides suitable initial conditions for test particle simulations exploring perturbations from the Galactic bar \citep[e.g.][]{dillamore2023}, satellites such as the LMC \citep[e.g.][]{garavito-camargo2019,conroy2021,erkal2021,chandra2025,brooks2025}, or figure rotation of the dark matter halo \citep{valluri2021}. For these purposes we make the fiducial potential and $N$-body stellar halo Schwarzschild model publicly available at \url{https://doi.org/10.5281/zenodo.17234304}.
\end{enumerate}

This study is a first attempt at building a Milky Way potential model which is consistent with the observed 3D structure of the stellar halo. However, we have made some simplifying assumptions, such as a static rigid dark matter halo with no radial dependence. Future studies will go further to build a fully self-consistent Milky Way halo model, including a live dark matter halo. This will enable a more accurate assessment of how tilted haloes evolve with time. We have also ignored perturbations from the LMC, which are causing the stellar halo to be pulled out of equilibrium \citep[e.g.][]{erkal2020,erkal2021,conroy2021}. In a subsequent paper we plan to update our constraints on the dark matter halo by taking these effects into account. Furthermore models can be fitted to new measurements of RR Lyrae in the stellar halo by \textit{Gaia} and Rubin-LSST, allowing its structure and dynamics to be revealed in unprecedented detail.

\section*{Acknowledgements}
We are grateful to the anonymous referee for a useful report. We thank Eugene Vasiliev, Vasily Belokurov and Hanyuan Zhang for helpful discussions during this study. AMD and JLS acknowledge support from the Royal Society (URF\textbackslash R1\textbackslash191555; URF\textbackslash R\textbackslash 241030).

\section*{Data Availability}
The code used in this project can be found at \url{https://github.com/adllmr/schwarzschild}. Our fiducial potential with a tilted prolate dark matter halo and an $N$-body snapshot of the corresponding GSE model are at \url{https://doi.org/10.5281/zenodo.17234304}. Other products from the model (e.g. full orbit integrations and weights) are available on request.



\bibliographystyle{mnras}
\bibliography{refs} 

\begin{thebibliography}{}
\makeatletter
\relax
\def\mn@urlcharsother{\let\do\@makeother \do\$\do\&\do\#\do\^\do\_\do\%\do\~}
\def\mn@doi{\begingroup\mn@urlcharsother \@ifnextchar [ {\mn@doi@} {\mn@doi@[]}}
\def\mn@doi@[#1]#2{\def\@tempa{#1}\ifx\@tempa\@empty \href {http://dx.doi.org/#2} {doi:#2}\else \href {http://dx.doi.org/#2} {#1}\fi \endgroup}
\def\mn@eprint#1#2{\mn@eprint@#1:#2::\@nil}
\def\mn@eprint@arXiv#1{\href {http://arxiv.org/abs/#1} {{\tt arXiv:#1}}}
\def\mn@eprint@dblp#1{\href {http://dblp.uni-trier.de/rec/bibtex/#1.xml} {dblp:#1}}
\def\mn@eprint@#1:#2:#3:#4\@nil{\def\@tempa {#1}\def\@tempb {#2}\def\@tempc {#3}\ifx \@tempc \@empty \let \@tempc \@tempb \let \@tempb \@tempa \fi \ifx \@tempb \@empty \def\@tempb {arXiv}\fi \@ifundefined {mn@eprint@\@tempb}{\@tempb:\@tempc}{\expandafter \expandafter \csname mn@eprint@\@tempb\endcsname \expandafter{\@tempc}}}

\bibitem[\protect\citeauthoryear{{An}, {Evans}  \& {Sanders}}{{An} et~al.}{2017}]{an2017}
{An} J.,  {Evans} N.~W.,   {Sanders} J.~L.,  2017, \mn@doi [\mnras] {10.1093/mnras/stx195}, \href {https://ui.adsabs.harvard.edu/abs/2017MNRAS.467.1281A} {467, 1281}

\bibitem[\protect\citeauthoryear{{Andersen}, {Dahl}  \& {Vandenberghe}}{{Andersen} et~al.}{2020}]{cvxopt}
{Andersen} M.,  {Dahl} J.,   {Vandenberghe} L.,  2020, {CVXOPT: Convex Optimization}, Astrophysics Source Code Library, record ascl:2008.017

\bibitem[\protect\citeauthoryear{{Balbinot} \& {Helmi}}{{Balbinot} \& {Helmi}}{2021}]{balbinot2021}
{Balbinot} E.,  {Helmi} A.,  2021, \mn@doi [\aap] {10.1051/0004-6361/202141015}, \href {https://ui.adsabs.harvard.edu/abs/2021A&A...654A..15B} {654, A15}

\bibitem[\protect\citeauthoryear{{Belokurov} et~al.,}{{Belokurov} et~al.}{2007}]{belokurov2007}
{Belokurov} V.,  et~al., 2007, \mn@doi [\apjl] {10.1086/513144}, \href {https://ui.adsabs.harvard.edu/abs/2007ApJ...657L..89B} {657, L89}

\bibitem[\protect\citeauthoryear{{Belokurov}, {Erkal}, {Evans}, {Koposov}  \& {Deason}}{{Belokurov} et~al.}{2018}]{belokurov2018}
{Belokurov} V.,  {Erkal} D.,  {Evans} N.~W.,  {Koposov} S.~E.,   {Deason} A.~J.,  2018, \mn@doi [\mnras] {10.1093/mnras/sty982}, \href {https://ui.adsabs.harvard.edu/abs/2018MNRAS.478..611B} {478, 611}

\bibitem[\protect\citeauthoryear{{Belokurov}, {Sanders}, {Fattahi}, {Smith}, {Deason}, {Evans}  \& {Grand}}{{Belokurov} et~al.}{2020}]{belokurov2020}
{Belokurov} V.,  {Sanders} J.~L.,  {Fattahi} A.,  {Smith} M.~C.,  {Deason} A.~J.,  {Evans} N.~W.,   {Grand} R. J.~J.,  2020, \mn@doi [\mnras] {10.1093/mnras/staa876}, \href {https://ui.adsabs.harvard.edu/abs/2020MNRAS.494.3880B} {494, 3880}

\bibitem[\protect\citeauthoryear{{Belokurov}, {Vasiliev}, {Deason}, {Koposov}, {Fattahi}, {Dillamore}, {Davies}  \& {Grand}}{{Belokurov} et~al.}{2023}]{belokurov_chevrons}
{Belokurov} V.,  {Vasiliev} E.,  {Deason} A.~J.,  {Koposov} S.~E.,  {Fattahi} A.,  {Dillamore} A.~M.,  {Davies} E.~Y.,   {Grand} R. J.~J.,  2023, \mn@doi [\mnras] {10.1093/mnras/stac3436}, \href {https://ui.adsabs.harvard.edu/abs/2023MNRAS.518.6200B} {518, 6200}

\bibitem[\protect\citeauthoryear{{Binney}}{{Binney}}{2012}]{binney2012}
{Binney} J.,  2012, \mn@doi [\mnras] {10.1111/j.1365-2966.2012.21757.x}, \href {https://ui.adsabs.harvard.edu/abs/2012MNRAS.426.1324B} {426, 1324}

\bibitem[\protect\citeauthoryear{{Binney} \& {Piffl}}{{Binney} \& {Piffl}}{2015}]{binney2015}
{Binney} J.,  {Piffl} T.,  2015, \mn@doi [\mnras] {10.1093/mnras/stv2225}, \href {https://ui.adsabs.harvard.edu/abs/2015MNRAS.454.3653B} {454, 3653}

\bibitem[\protect\citeauthoryear{{Binney} \& {Tremaine}}{{Binney} \& {Tremaine}}{2008}]{binney_tremaine}
{Binney} J.,  {Tremaine} S.,  2008, {Galactic Dynamics: Second Edition}

\bibitem[\protect\citeauthoryear{{Bird}, {Xue}, {Liu}, {Shen}, {Flynn}, {Yang}, {Zhao}  \& {Tian}}{{Bird} et~al.}{2021}]{bird2021}
{Bird} S.~A.,  {Xue} X.-X.,  {Liu} C.,  {Shen} J.,  {Flynn} C.,  {Yang} C.,  {Zhao} G.,   {Tian} H.-J.,  2021, \mn@doi [\apj] {10.3847/1538-4357/abfa9e}, \href {https://ui.adsabs.harvard.edu/abs/2021ApJ...919...66B} {919, 66}

\bibitem[\protect\citeauthoryear{{Bland-Hawthorn} \& {Gerhard}}{{Bland-Hawthorn} \& {Gerhard}}{2016}]{bland-hawthorn2016}
{Bland-Hawthorn} J.,  {Gerhard} O.,  2016, \mn@doi [\araa] {10.1146/annurev-astro-081915-023441}, \href {https://ui.adsabs.harvard.edu/abs/2016ARA&A..54..529B} {54, 529}

\bibitem[\protect\citeauthoryear{{Bonaca} et~al.,}{{Bonaca} et~al.}{2020}]{bonaca2020}
{Bonaca} A.,  et~al., 2020, \mn@doi [\apjl] {10.3847/2041-8213/ab9caa}, \href {https://ui.adsabs.harvard.edu/abs/2020ApJ...897L..18B} {897, L18}

\bibitem[\protect\citeauthoryear{{Bovy}, {Bahmanyar}, {Fritz}  \& {Kallivayalil}}{{Bovy} et~al.}{2016}]{bovy2016}
{Bovy} J.,  {Bahmanyar} A.,  {Fritz} T.~K.,   {Kallivayalil} N.,  2016, \mn@doi [\apj] {10.3847/1538-4357/833/1/31}, \href {https://ui.adsabs.harvard.edu/abs/2016ApJ...833...31B} {833, 31}

\bibitem[\protect\citeauthoryear{{Bowden}, {Belokurov}  \& {Evans}}{{Bowden} et~al.}{2015}]{bowden2014}
{Bowden} A.,  {Belokurov} V.,   {Evans} N.~W.,  2015, \mn@doi [\mnras] {10.1093/mnras/stv285}, \href {https://ui.adsabs.harvard.edu/abs/2015MNRAS.449.1391B} {449, 1391}

\bibitem[\protect\citeauthoryear{{Brooks}, {Sanders}, {Dillamore}, {Garavito-Camargo}  \& {Price-Whelan}}{{Brooks} et~al.}{2026}]{brooks2025}
{Brooks} R. A.~N.,  {Sanders} J.~L.,  {Dillamore} A.~M.,  {Garavito-Camargo} N.,   {Price-Whelan} A.~M.,  2026, \mn@doi [\mnras] {10.1093/mnras/staf2111}, \href {https://ui.adsabs.harvard.edu/abs/2026MNRAS.545f2111B} {545, staf2111}

\bibitem[\protect\citeauthoryear{{Chandra} et~al.,}{{Chandra} et~al.}{2025}]{chandra2025}
{Chandra} V.,  et~al., 2025, \mn@doi [\apj] {10.3847/1538-4357/addab6}, \href {https://ui.adsabs.harvard.edu/abs/2025ApJ...988..156C} {988, 156}

\bibitem[\protect\citeauthoryear{{Chua}, {Pillepich}, {Vogelsberger}  \& {Hernquist}}{{Chua} et~al.}{2019}]{chua2019}
{Chua} K. T.~E.,  {Pillepich} A.,  {Vogelsberger} M.,   {Hernquist} L.,  2019, \mn@doi [\mnras] {10.1093/mnras/sty3531}, \href {https://ui.adsabs.harvard.edu/abs/2019MNRAS.484..476C} {484, 476}

\bibitem[\protect\citeauthoryear{{Conroy} et~al.,}{{Conroy} et~al.}{2019}]{conroy2019}
{Conroy} C.,  et~al., 2019, \mn@doi [\apj] {10.3847/1538-4357/ab38b8}, \href {https://ui.adsabs.harvard.edu/abs/2019ApJ...883..107C} {883, 107}

\bibitem[\protect\citeauthoryear{{Conroy}, {Naidu}, {Garavito-Camargo}, {Besla}, {Zaritsky}, {Bonaca}  \& {Johnson}}{{Conroy} et~al.}{2021}]{conroy2021}
{Conroy} C.,  {Naidu} R.~P.,  {Garavito-Camargo} N.,  {Besla} G.,  {Zaritsky} D.,  {Bonaca} A.,   {Johnson} B.~D.,  2021, \mn@doi [\nat] {10.1038/s41586-021-03385-7}, \href {https://ui.adsabs.harvard.edu/abs/2021Natur.592..534C} {592, 534}

\bibitem[\protect\citeauthoryear{{Cunningham} et~al.,}{{Cunningham} et~al.}{2019}]{cunningham2019}
{Cunningham} E.~C.,  et~al., 2019, \mn@doi [\apj] {10.3847/1538-4357/ab24cd}, \href {https://ui.adsabs.harvard.edu/abs/2019ApJ...879..120C} {879, 120}

\bibitem[\protect\citeauthoryear{{Das} \& {Binney}}{{Das} \& {Binney}}{2016}]{das2016}
{Das} P.,  {Binney} J.,  2016, \mn@doi [\mnras] {10.1093/mnras/stw744}, \href {https://ui.adsabs.harvard.edu/abs/2016MNRAS.460.1725D} {460, 1725}

\bibitem[\protect\citeauthoryear{{Di Matteo}, {Haywood}, {Lehnert}, {Katz}, {Khoperskov}, {Snaith}, {G{\'o}mez}  \& {Robichon}}{{Di Matteo} et~al.}{2019}]{dimatteo2019}
{Di Matteo} P.,  {Haywood} M.,  {Lehnert} M.~D.,  {Katz} D.,  {Khoperskov} S.,  {Snaith} O.~N.,  {G{\'o}mez} A.,   {Robichon} N.,  2019, \mn@doi [\aap] {10.1051/0004-6361/201834929}, \href {https://ui.adsabs.harvard.edu/abs/2019A&A...632A...4D} {632, A4}

\bibitem[\protect\citeauthoryear{{Dillamore} \& {Sanders}}{{Dillamore} \& {Sanders}}{2025}]{dillamore2025b}
{Dillamore} A.~M.,  {Sanders} J.~L.,  2025, \mn@doi [\mnras] {10.1093/mnras/staf1264}, \href {https://ui.adsabs.harvard.edu/abs/2025MNRAS.542.1331D} {542, 1331}

\bibitem[\protect\citeauthoryear{{Dillamore}, {Belokurov}, {Font}  \& {McCarthy}}{{Dillamore} et~al.}{2022}]{dillamore2022}
{Dillamore} A.~M.,  {Belokurov} V.,  {Font} A.~S.,   {McCarthy} I.~G.,  2022, \mn@doi [\mnras] {10.1093/mnras/stac1038}, \href {https://ui.adsabs.harvard.edu/abs/2022MNRAS.513.1867D} {513, 1867}

\bibitem[\protect\citeauthoryear{{Dillamore}, {Belokurov}, {Evans}  \& {Davies}}{{Dillamore} et~al.}{2023}]{dillamore2023}
{Dillamore} A.~M.,  {Belokurov} V.,  {Evans} N.~W.,   {Davies} E.~Y.,  2023, \mn@doi [\mnras] {10.1093/mnras/stad2136}, \href {https://ui.adsabs.harvard.edu/abs/2023MNRAS.524.3596D} {524, 3596}

\bibitem[\protect\citeauthoryear{{Dillamore}, {Belokurov}  \& {Evans}}{{Dillamore} et~al.}{2024}]{dillamore2024}
{Dillamore} A.~M.,  {Belokurov} V.,   {Evans} N.~W.,  2024, \mn@doi [\mnras] {10.1093/mnras/stae1789}, \href {https://ui.adsabs.harvard.edu/abs/2024MNRAS.532.4389D} {532, 4389}

\bibitem[\protect\citeauthoryear{{Dodd}, {Helmi}  \& {Koppelman}}{{Dodd} et~al.}{2022}]{dodd2022}
{Dodd} E.,  {Helmi} A.,   {Koppelman} H.~H.,  2022, \mn@doi [\aap] {10.1051/0004-6361/202141354}, \href {https://ui.adsabs.harvard.edu/abs/2022A&A...659A..61D} {659, A61}

\bibitem[\protect\citeauthoryear{{Donlon}, {Newberg}, {Weiss}, {Amy}  \& {Thompson}}{{Donlon} et~al.}{2019}]{donlon2019}
{Donlon} II T.,  {Newberg} H.~J.,  {Weiss} J.,  {Amy} P.,   {Thompson} J.,  2019, \mn@doi [\apj] {10.3847/1538-4357/ab4f72}, \href {https://ui.adsabs.harvard.edu/abs/2019ApJ...886...76D} {886, 76}

\bibitem[\protect\citeauthoryear{{Donlon}, {Newberg}, {Sanderson}, {Bregou}, {Horta}, {Arora}  \& {Panithanpaisal}}{{Donlon} et~al.}{2024}]{donlon2023}
{Donlon} T.,  {Newberg} H.~J.,  {Sanderson} R.,  {Bregou} E.,  {Horta} D.,  {Arora} A.,   {Panithanpaisal} N.,  2024, \mn@doi [\mnras] {10.1093/mnras/stae1264}, \href {https://ui.adsabs.harvard.edu/abs/2024MNRAS.531.1422D} {531, 1422}

\bibitem[\protect\citeauthoryear{{Drakos}, {Taylor}, {Berrouet}, {Robotham}  \& {Power}}{{Drakos} et~al.}{2019}]{drakos2019}
{Drakos} N.~E.,  {Taylor} J.~E.,  {Berrouet} A.,  {Robotham} A. S.~G.,   {Power} C.,  2019, \mn@doi [\mnras] {10.1093/mnras/stz1306}, \href {https://ui.adsabs.harvard.edu/abs/2019MNRAS.487..993D} {487, 993}

\bibitem[\protect\citeauthoryear{{Emami} et~al.,}{{Emami} et~al.}{2021}]{emami2021}
{Emami} R.,  et~al., 2021, \mn@doi [\apj] {10.3847/1538-4357/abf147}, \href {https://ui.adsabs.harvard.edu/abs/2021ApJ...913...36E} {913, 36}

\bibitem[\protect\citeauthoryear{{Erkal}, {Sanders}  \& {Belokurov}}{{Erkal} et~al.}{2016}]{erkal2016}
{Erkal} D.,  {Sanders} J.~L.,   {Belokurov} V.,  2016, \mn@doi [\mnras] {10.1093/mnras/stw1400}, \href {https://ui.adsabs.harvard.edu/abs/2016MNRAS.461.1590E} {461, 1590}

\bibitem[\protect\citeauthoryear{{Erkal} et~al.,}{{Erkal} et~al.}{2019}]{erkal2019}
{Erkal} D.,  et~al., 2019, \mn@doi [\mnras] {10.1093/mnras/stz1371}, \href {https://ui.adsabs.harvard.edu/abs/2019MNRAS.487.2685E} {487, 2685}

\bibitem[\protect\citeauthoryear{{Erkal}, {Belokurov}  \& {Parkin}}{{Erkal} et~al.}{2020}]{erkal2020}
{Erkal} D.,  {Belokurov} V.~A.,   {Parkin} D.~L.,  2020, \mn@doi [\mnras] {10.1093/mnras/staa2840}, \href {https://ui.adsabs.harvard.edu/abs/2020MNRAS.498.5574E} {498, 5574}

\bibitem[\protect\citeauthoryear{{Erkal} et~al.,}{{Erkal} et~al.}{2021}]{erkal2021}
{Erkal} D.,  et~al., 2021, \mn@doi [\mnras] {10.1093/mnras/stab1828}, \href {https://ui.adsabs.harvard.edu/abs/2021MNRAS.506.2677E} {506, 2677}

\bibitem[\protect\citeauthoryear{{Fattahi} et~al.,}{{Fattahi} et~al.}{2019}]{fattahi2019}
{Fattahi} A.,  et~al., 2019, \mn@doi [\mnras] {10.1093/mnras/stz159}, \href {https://ui.adsabs.harvard.edu/abs/2019MNRAS.484.4471F} {484, 4471}

\bibitem[\protect\citeauthoryear{{Gaia Collaboration} et~al.,}{{Gaia Collaboration} et~al.}{2016}]{gaia}
{Gaia Collaboration} et~al., 2016, \mn@doi [\aap] {10.1051/0004-6361/201629272}, \href {https://ui.adsabs.harvard.edu/abs/2016A&A...595A...1G} {595, A1}

\bibitem[\protect\citeauthoryear{{Gaia Collaboration} et~al.,}{{Gaia Collaboration} et~al.}{2023}]{gaia_dr3}
{Gaia Collaboration} et~al., 2023, \mn@doi [\aap] {10.1051/0004-6361/202243940}, \href {https://ui.adsabs.harvard.edu/abs/2023A&A...674A...1G} {674, A1}

\bibitem[\protect\citeauthoryear{{Gallart}, {Bernard}, {Brook}, {Ruiz-Lara}, {Cassisi}, {Hill}  \& {Monelli}}{{Gallart} et~al.}{2019}]{gallart2019}
{Gallart} C.,  {Bernard} E.~J.,  {Brook} C.~B.,  {Ruiz-Lara} T.,  {Cassisi} S.,  {Hill} V.,   {Monelli} M.,  2019, \mn@doi [Nature Astronomy] {10.1038/s41550-019-0829-5}, \href {https://ui.adsabs.harvard.edu/abs/2019NatAs...3..932G} {3, 932}

\bibitem[\protect\citeauthoryear{{Garavito-Camargo}, {Besla}, {Laporte}, {Johnston}, {G{\'o}mez}  \& {Watkins}}{{Garavito-Camargo} et~al.}{2019}]{garavito-camargo2019}
{Garavito-Camargo} N.,  {Besla} G.,  {Laporte} C. F.~P.,  {Johnston} K.~V.,  {G{\'o}mez} F.~A.,   {Watkins} L.~L.,  2019, \mn@doi [\apj] {10.3847/1538-4357/ab32eb}, \href {https://ui.adsabs.harvard.edu/abs/2019ApJ...884...51G} {884, 51}

\bibitem[\protect\citeauthoryear{{Han} et~al.,}{{Han} et~al.}{2022a}]{han2022}
{Han} J.~J.,  et~al., 2022a, \mn@doi [\aj] {10.3847/1538-3881/ac97e9}, \href {https://ui.adsabs.harvard.edu/abs/2022AJ....164..249H} {164, 249}

\bibitem[\protect\citeauthoryear{{Han} et~al.,}{{Han} et~al.}{2022b}]{han2022b}
{Han} J.~J.,  et~al., 2022b, \mn@doi [\apj] {10.3847/1538-4357/ac795f}, \href {https://ui.adsabs.harvard.edu/abs/2022ApJ...934...14H} {934, 14}

\bibitem[\protect\citeauthoryear{{Han}, {Conroy}  \& {Hernquist}}{{Han} et~al.}{2023a}]{han2023a}
{Han} J.~J.,  {Conroy} C.,   {Hernquist} L.,  2023a, \mn@doi [Nature Astronomy] {10.1038/s41550-023-02076-9}, \href {https://ui.adsabs.harvard.edu/abs/2023NatAs...7.1481H} {7, 1481}

\bibitem[\protect\citeauthoryear{{Han}, {Semenov}, {Conroy}  \& {Hernquist}}{{Han} et~al.}{2023b}]{han2023}
{Han} J.~J.,  {Semenov} V.,  {Conroy} C.,   {Hernquist} L.,  2023b, \mn@doi [\apjl] {10.3847/2041-8213/ad0641}, \href {https://ui.adsabs.harvard.edu/abs/2023ApJ...957L..24H} {957, L24}

\bibitem[\protect\citeauthoryear{{Han}, {Conroy}, {Zaritsky}, {Bonaca}, {Caldwell}, {Chandra}  \& {Ting}}{{Han} et~al.}{2024}]{han2024}
{Han} J.~J.,  {Conroy} C.,  {Zaritsky} D.,  {Bonaca} A.,  {Caldwell} N.,  {Chandra} V.,   {Ting} Y.-S.,  2024, \mn@doi [arXiv e-prints] {10.48550/arXiv.2406.12969}, \href {https://ui.adsabs.harvard.edu/abs/2024arXiv240612969H} {p. arXiv:2406.12969}

\bibitem[\protect\citeauthoryear{{Hattori}, {Valluri}  \& {Vasiliev}}{{Hattori} et~al.}{2021}]{hattori2021}
{Hattori} K.,  {Valluri} M.,   {Vasiliev} E.,  2021, \mn@doi [\mnras] {10.1093/mnras/stab2898}, \href {https://ui.adsabs.harvard.edu/abs/2021MNRAS.508.5468H} {508, 5468}

\bibitem[\protect\citeauthoryear{{Helmi}, {Babusiaux}, {Koppelman}, {Massari}, {Veljanoski}  \& {Brown}}{{Helmi} et~al.}{2018}]{helmi2018}
{Helmi} A.,  {Babusiaux} C.,  {Koppelman} H.~H.,  {Massari} D.,  {Veljanoski} J.,   {Brown} A. G.~A.,  2018, \mn@doi [\nat] {10.1038/s41586-018-0625-x}, \href {https://ui.adsabs.harvard.edu/abs/2018Natur.563...85H} {563, 85}

\bibitem[\protect\citeauthoryear{{Horta}, {Lu}, {Ness}, {Lisanti}  \& {Price-Whelan}}{{Horta} et~al.}{2024}]{horta2024_gse}
{Horta} D.,  {Lu} Y.~L.,  {Ness} M.~K.,  {Lisanti} M.,   {Price-Whelan} A.~M.,  2024, \mn@doi [\apj] {10.3847/1538-4357/ad58de}, \href {https://ui.adsabs.harvard.edu/abs/2024ApJ...971..170H} {971, 170}

\bibitem[\protect\citeauthoryear{{Hunt} \& {Vasiliev}}{{Hunt} \& {Vasiliev}}{2025}]{hunt2025}
{Hunt} J. A.~S.,  {Vasiliev} E.,  2025, \mn@doi [\nar] {10.1016/j.newar.2024.101721}, \href {https://ui.adsabs.harvard.edu/abs/2025NewAR.10001721H} {100, 101721}

\bibitem[\protect\citeauthoryear{{Ibata} et~al.,}{{Ibata} et~al.}{2024}]{ibata2024}
{Ibata} R.,  et~al., 2024, \mn@doi [\apj] {10.3847/1538-4357/ad382d}, \href {https://ui.adsabs.harvard.edu/abs/2024ApJ...967...89I} {967, 89}

\bibitem[\protect\citeauthoryear{{Iorio} \& {Belokurov}}{{Iorio} \& {Belokurov}}{2019}]{iorio2019}
{Iorio} G.,  {Belokurov} V.,  2019, \mn@doi [\mnras] {10.1093/mnras/sty2806}, \href {https://ui.adsabs.harvard.edu/abs/2019MNRAS.482.3868I} {482, 3868}

\bibitem[\protect\citeauthoryear{{Iorio} \& {Belokurov}}{{Iorio} \& {Belokurov}}{2021}]{iorio2021}
{Iorio} G.,  {Belokurov} V.,  2021, \mn@doi [\mnras] {10.1093/mnras/stab005}, \href {https://ui.adsabs.harvard.edu/abs/2021MNRAS.502.5686I} {502, 5686}

\bibitem[\protect\citeauthoryear{{Iorio}, {Belokurov}, {Erkal}, {Koposov}, {Nipoti}  \& {Fraternali}}{{Iorio} et~al.}{2018}]{iorio2018}
{Iorio} G.,  {Belokurov} V.,  {Erkal} D.,  {Koposov} S.~E.,  {Nipoti} C.,   {Fraternali} F.,  2018, \mn@doi [\mnras] {10.1093/mnras/stx2819}, \href {https://ui.adsabs.harvard.edu/abs/2018MNRAS.474.2142I} {474, 2142}

\bibitem[\protect\citeauthoryear{{Juri{\'c}} et~al.,}{{Juri{\'c}} et~al.}{2008}]{juric2008}
{Juri{\'c}} M.,  et~al., 2008, \mn@doi [\apj] {10.1086/523619}, \href {https://ui.adsabs.harvard.edu/abs/2008ApJ...673..864J} {673, 864}

\bibitem[\protect\citeauthoryear{{Koposov}, {Rix}  \& {Hogg}}{{Koposov} et~al.}{2010}]{koposov2010}
{Koposov} S.~E.,  {Rix} H.-W.,   {Hogg} D.~W.,  2010, \mn@doi [\apj] {10.1088/0004-637X/712/1/260}, \href {https://ui.adsabs.harvard.edu/abs/2010ApJ...712..260K} {712, 260}

\bibitem[\protect\citeauthoryear{{Koposov} et~al.,}{{Koposov} et~al.}{2023}]{koposov2023}
{Koposov} S.~E.,  et~al., 2023, \mn@doi [\mnras] {10.1093/mnras/stad551}, \href {https://ui.adsabs.harvard.edu/abs/2023MNRAS.521.4936K} {521, 4936}

\bibitem[\protect\citeauthoryear{{K{\"u}pper}, {Balbinot}, {Bonaca}, {Johnston}, {Hogg}, {Kroupa}  \& {Santiago}}{{K{\"u}pper} et~al.}{2015}]{kupper2015}
{K{\"u}pper} A. H.~W.,  {Balbinot} E.,  {Bonaca} A.,  {Johnston} K.~V.,  {Hogg} D.~W.,  {Kroupa} P.,   {Santiago} B.~X.,  2015, \mn@doi [\apj] {10.1088/0004-637X/803/2/80}, \href {https://ui.adsabs.harvard.edu/abs/2015ApJ...803...80K} {803, 80}

\bibitem[\protect\citeauthoryear{{Lancaster}, {Koposov}, {Belokurov}, {Evans}  \& {Deason}}{{Lancaster} et~al.}{2019}]{lancaster2019}
{Lancaster} L.,  {Koposov} S.~E.,  {Belokurov} V.,  {Evans} N.~W.,   {Deason} A.~J.,  2019, \mn@doi [\mnras] {10.1093/mnras/stz853}, \href {https://ui.adsabs.harvard.edu/abs/2019MNRAS.486..378L} {486, 378}

\bibitem[\protect\citeauthoryear{{Lane} \& {Bovy}}{{Lane} \& {Bovy}}{2025}]{lane2025}
{Lane} J. M.~M.,  {Bovy} J.,  2025, arXiv e-prints, \href {https://ui.adsabs.harvard.edu/abs/2025arXiv250904557L} {p. arXiv:2509.04557}

\bibitem[\protect\citeauthoryear{{Lane}, {Bovy}  \& {Mackereth}}{{Lane} et~al.}{2023}]{lane2023}
{Lane} J. M.~M.,  {Bovy} J.,   {Mackereth} J.~T.,  2023, \mn@doi [\mnras] {10.1093/mnras/stad2834}, \href {https://ui.adsabs.harvard.edu/abs/2023MNRAS.526.1209L} {526, 1209}

\bibitem[\protect\citeauthoryear{{Law} \& {Majewski}}{{Law} \& {Majewski}}{2010}]{law2010}
{Law} D.~R.,  {Majewski} S.~R.,  2010, \mn@doi [\apj] {10.1088/0004-637X/714/1/229}, \href {https://ui.adsabs.harvard.edu/abs/2010ApJ...714..229L} {714, 229}

\bibitem[\protect\citeauthoryear{{Lilleengen} et~al.,}{{Lilleengen} et~al.}{2023}]{lilleengen2023}
{Lilleengen} S.,  et~al., 2023, \mn@doi [\mnras] {10.1093/mnras/stac3108}, \href {https://ui.adsabs.harvard.edu/abs/2023MNRAS.518..774L} {518, 774}

\bibitem[\protect\citeauthoryear{{Mackereth} \& {Bovy}}{{Mackereth} \& {Bovy}}{2020}]{mackereth2020}
{Mackereth} J.~T.,  {Bovy} J.,  2020, \mn@doi [\mnras] {10.1093/mnras/staa047}, \href {https://ui.adsabs.harvard.edu/abs/2020MNRAS.492.3631M} {492, 3631}

\bibitem[\protect\citeauthoryear{{Malhan} \& {Ibata}}{{Malhan} \& {Ibata}}{2019}]{malhan2019}
{Malhan} K.,  {Ibata} R.~A.,  2019, \mn@doi [\mnras] {10.1093/mnras/stz1035}, \href {https://ui.adsabs.harvard.edu/abs/2019MNRAS.486.2995M} {486, 2995}

\bibitem[\protect\citeauthoryear{{McMillan}}{{McMillan}}{2017}]{mcmillan17}
{McMillan} P.~J.,  2017, \mn@doi [\mnras] {10.1093/mnras/stw2759}, \href {https://ui.adsabs.harvard.edu/abs/2017MNRAS.465...76M} {465, 76}

\bibitem[\protect\citeauthoryear{{Merritt}}{{Merritt}}{1996}]{merritt1996}
{Merritt} D.,  1996, \mn@doi [Celestial Mechanics and Dynamical Astronomy] {10.1007/BF00051605}, \href {https://ui.adsabs.harvard.edu/abs/1996CeMDA..64...55M} {64, 55}

\bibitem[\protect\citeauthoryear{{Naidu} et~al.,}{{Naidu} et~al.}{2021}]{naidu2021}
{Naidu} R.~P.,  et~al., 2021, \mn@doi [\apj] {10.3847/1538-4357/ac2d2d}, \href {https://ui.adsabs.harvard.edu/abs/2021ApJ...923...92N} {923, 92}

\bibitem[\protect\citeauthoryear{{Navarro}, {Frenk}  \& {White}}{{Navarro} et~al.}{1997}]{NFW}
{Navarro} J.~F.,  {Frenk} C.~S.,   {White} S. D.~M.,  1997, \mn@doi [\apj] {10.1086/304888}, \href {https://ui.adsabs.harvard.edu/abs/1997ApJ...490..493N} {490, 493}

\bibitem[\protect\citeauthoryear{{Nelson} et~al.,}{{Nelson} et~al.}{2015}]{illustris}
{Nelson} D.,  et~al., 2015, \mn@doi [Astronomy and Computing] {10.1016/j.ascom.2015.09.003}, \href {https://ui.adsabs.harvard.edu/abs/2015A&C....13...12N} {13, 12}

\bibitem[\protect\citeauthoryear{{Nelson} et~al.,}{{Nelson} et~al.}{2019a}]{nelson2019}
{Nelson} D.,  et~al., 2019a, \mn@doi [Computational Astrophysics and Cosmology] {10.1186/s40668-019-0028-x}, \href {https://ui.adsabs.harvard.edu/abs/2019ComAC...6....2N} {6, 2}

\bibitem[\protect\citeauthoryear{{Nelson} et~al.,}{{Nelson} et~al.}{2019b}]{tng50}
{Nelson} D.,  et~al., 2019b, \mn@doi [\mnras] {10.1093/mnras/stz2306}, \href {https://ui.adsabs.harvard.edu/abs/2019MNRAS.490.3234N} {490, 3234}

\bibitem[\protect\citeauthoryear{{Nibauer} \& {Bonaca}}{{Nibauer} \& {Bonaca}}{2025}]{nibauer2025}
{Nibauer} J.,  {Bonaca} A.,  2025, \mn@doi [\apjl] {10.3847/2041-8213/add0a9}, \href {https://ui.adsabs.harvard.edu/abs/2025ApJ...985L..22N} {985, L22}

\bibitem[\protect\citeauthoryear{{Perottoni}, {Limberg}, {Amarante}, {Rossi}, {Queiroz}, {Santucci}, {P{\'e}rez-Villegas}  \& {Chiappini}}{{Perottoni} et~al.}{2022}]{perottoni2022}
{Perottoni} H.~D.,  {Limberg} G.,  {Amarante} J. A.~S.,  {Rossi} S.,  {Queiroz} A. B.~A.,  {Santucci} R.~M.,  {P{\'e}rez-Villegas} A.,   {Chiappini} C.,  2022, \mn@doi [\apjl] {10.3847/2041-8213/ac88d6}, \href {https://ui.adsabs.harvard.edu/abs/2022ApJ...936L...2P} {936, L2}

\bibitem[\protect\citeauthoryear{{Piffl}, {Penoyre}  \& {Binney}}{{Piffl} et~al.}{2015}]{piffl2015}
{Piffl} T.,  {Penoyre} Z.,   {Binney} J.,  2015, \mn@doi [\mnras] {10.1093/mnras/stv938}, \href {https://ui.adsabs.harvard.edu/abs/2015MNRAS.451..639P} {451, 639}

\bibitem[\protect\citeauthoryear{{Pillepich} et~al.,}{{Pillepich} et~al.}{2019}]{pillepich2019}
{Pillepich} A.,  et~al., 2019, \mn@doi [\mnras] {10.1093/mnras/stz2338}, \href {https://ui.adsabs.harvard.edu/abs/2019MNRAS.490.3196P} {490, 3196}

\bibitem[\protect\citeauthoryear{{Pillepich} et~al.,}{{Pillepich} et~al.}{2024}]{pillepich2024}
{Pillepich} A.,  et~al., 2024, \mn@doi [\mnras] {10.1093/mnras/stae2165}, \href {https://ui.adsabs.harvard.edu/abs/2024MNRAS.535.1721P} {535, 1721}

\bibitem[\protect\citeauthoryear{{Posti} \& {Helmi}}{{Posti} \& {Helmi}}{2019}]{posti2019}
{Posti} L.,  {Helmi} A.,  2019, \mn@doi [\aap] {10.1051/0004-6361/201833355}, \href {https://ui.adsabs.harvard.edu/abs/2019A&A...621A..56P} {621, A56}

\bibitem[\protect\citeauthoryear{{Sanders} \& {Binney}}{{Sanders} \& {Binney}}{2014}]{sanders2014}
{Sanders} J.~L.,  {Binney} J.,  2014, \mn@doi [\mnras] {10.1093/mnras/stu796}, \href {https://ui.adsabs.harvard.edu/abs/2014MNRAS.441.3284S} {441, 3284}

\bibitem[\protect\citeauthoryear{{Sanders} \& {Evans}}{{Sanders} \& {Evans}}{2015}]{sanders2015}
{Sanders} J.~L.,  {Evans} N.~W.,  2015, \mn@doi [\mnras] {10.1093/mnras/stv1898}, \href {https://ui.adsabs.harvard.edu/abs/2015MNRAS.454..299S} {454, 299}

\bibitem[\protect\citeauthoryear{{Schwarzschild}}{{Schwarzschild}}{1979}]{schwarzschild1979}
{Schwarzschild} M.,  1979, \mn@doi [\apj] {10.1086/157282}, \href {https://ui.adsabs.harvard.edu/abs/1979ApJ...232..236S} {232, 236}

\bibitem[\protect\citeauthoryear{{Schwarzschild}}{{Schwarzschild}}{1982}]{schwarzschild1982}
{Schwarzschild} M.,  1982, \mn@doi [\apj] {10.1086/160531}, \href {https://ui.adsabs.harvard.edu/abs/1982ApJ...263..599S} {263, 599}

\bibitem[\protect\citeauthoryear{{Simion}, {Belokurov}  \& {Koposov}}{{Simion} et~al.}{2019}]{simion2019}
{Simion} I.~T.,  {Belokurov} V.,   {Koposov} S.~E.,  2019, \mn@doi [\mnras] {10.1093/mnras/sty2744}, \href {https://ui.adsabs.harvard.edu/abs/2019MNRAS.482..921S} {482, 921}

\bibitem[\protect\citeauthoryear{{Tenneti}, {Mandelbaum}  \& {Di Matteo}}{{Tenneti} et~al.}{2016}]{tenneti2016}
{Tenneti} A.,  {Mandelbaum} R.,   {Di Matteo} T.,  2016, \mn@doi [\mnras] {10.1093/mnras/stw1823}, \href {https://ui.adsabs.harvard.edu/abs/2016MNRAS.462.2668T} {462, 2668}

\bibitem[\protect\citeauthoryear{{Valluri}, {Price-Whelan}  \& {Snyder}}{{Valluri} et~al.}{2021}]{valluri2021}
{Valluri} M.,  {Price-Whelan} A.~M.,   {Snyder} S.~J.,  2021, \mn@doi [\apj] {10.3847/1538-4357/abe534}, \href {https://ui.adsabs.harvard.edu/abs/2021ApJ...910..150V} {910, 150}

\bibitem[\protect\citeauthoryear{{Vasiliev}}{{Vasiliev}}{2013}]{vasiliev2013}
{Vasiliev} E.,  2013, \mn@doi [\mnras] {10.1093/mnras/stt1235}, \href {https://ui.adsabs.harvard.edu/abs/2013MNRAS.434.3174V} {434, 3174}

\bibitem[\protect\citeauthoryear{{Vasiliev}}{{Vasiliev}}{2019}]{agama}
{Vasiliev} E.,  2019, \mn@doi [\mnras] {10.1093/mnras/sty2672}, \href {https://ui.adsabs.harvard.edu/abs/2019MNRAS.482.1525V} {482, 1525}

\bibitem[\protect\citeauthoryear{{Vasiliev} \& {Valluri}}{{Vasiliev} \& {Valluri}}{2020}]{vasiliev2020}
{Vasiliev} E.,  {Valluri} M.,  2020, \mn@doi [\apj] {10.3847/1538-4357/ab5fe0}, \href {https://ui.adsabs.harvard.edu/abs/2020ApJ...889...39V} {889, 39}

\bibitem[\protect\citeauthoryear{{Vasiliev}, {Belokurov}  \& {Erkal}}{{Vasiliev} et~al.}{2021}]{vasiliev_tango}
{Vasiliev} E.,  {Belokurov} V.,   {Erkal} D.,  2021, \mn@doi [\mnras] {10.1093/mnras/staa3673}, \href {https://ui.adsabs.harvard.edu/abs/2021MNRAS.501.2279V} {501, 2279}

\bibitem[\protect\citeauthoryear{{Vivas} et~al.,}{{Vivas} et~al.}{2001}]{vivas2001}
{Vivas} A.~K.,  et~al., 2001, \mn@doi [\apjl] {10.1086/320915}, \href {https://ui.adsabs.harvard.edu/abs/2001ApJ...554L..33V} {554, L33}

\bibitem[\protect\citeauthoryear{{Wang}, {Hammer}  \& {Yang}}{{Wang} et~al.}{2022}]{wang2022}
{Wang} J.,  {Hammer} F.,   {Yang} Y.,  2022, \mn@doi [\mnras] {10.1093/mnras/stab3258}, \href {https://ui.adsabs.harvard.edu/abs/2022MNRAS.510.2242W} {510, 2242}

\bibitem[\protect\citeauthoryear{{Woudenberg} \& {Helmi}}{{Woudenberg} \& {Helmi}}{2024}]{woudenberg2024}
{Woudenberg} H.~C.,  {Helmi} A.,  2024, \mn@doi [\aap] {10.1051/0004-6361/202451743}, \href {https://ui.adsabs.harvard.edu/abs/2024A&A...691A.277W} {691, A277}

\bibitem[\protect\citeauthoryear{{Woudenberg} \& {Helmi}}{{Woudenberg} \& {Helmi}}{2025}]{woudenberg2025}
{Woudenberg} H.~C.,  {Helmi} A.,  2025, \mn@doi [\aap] {10.1051/0004-6361/202555672}, \href {https://ui.adsabs.harvard.edu/abs/2025A&A...700A.240W} {700, A240}

\bibitem[\protect\citeauthoryear{{Wu}, {Zhao}, {Xue}, {Pei}  \& {Yang}}{{Wu} et~al.}{2022}]{wu2022}
{Wu} W.,  {Zhao} G.,  {Xue} X.-X.,  {Pei} W.,   {Yang} C.,  2022, \mn@doi [\aj] {10.3847/1538-3881/ac746e}, \href {https://ui.adsabs.harvard.edu/abs/2022AJ....164...41W} {164, 41}

\bibitem[\protect\citeauthoryear{{Yavetz}, {Johnston}, {Pearson}, {Price-Whelan}  \& {Weinberg}}{{Yavetz} et~al.}{2021}]{yavetz2021}
{Yavetz} T.~D.,  {Johnston} K.~V.,  {Pearson} S.,  {Price-Whelan} A.~M.,   {Weinberg} M.~D.,  2021, \mn@doi [\mnras] {10.1093/mnras/staa3687}, \href {https://ui.adsabs.harvard.edu/abs/2021MNRAS.501.1791Y} {501, 1791}

\bibitem[\protect\citeauthoryear{{Yavetz}, {Johnston}, {Pearson}, {Price-Whelan}  \& {Hamilton}}{{Yavetz} et~al.}{2023}]{yavetz2023}
{Yavetz} T.~D.,  {Johnston} K.~V.,  {Pearson} S.,  {Price-Whelan} A.~M.,   {Hamilton} C.,  2023, \mn@doi [\apj] {10.3847/1538-4357/ace7b9}, \href {https://ui.adsabs.harvard.edu/abs/2023ApJ...954..215Y} {954, 215}

\bibitem[\protect\citeauthoryear{{Ye}, {Du}, {Deng}, {Liao}, {Huang}, {Shi}  \& {Ma}}{{Ye} et~al.}{2024}]{ye2024}
{Ye} D.,  {Du} C.,  {Deng} M.,  {Liao} J.,  {Huang} Y.,  {Shi} J.,   {Ma} J.,  2024, \mn@doi [\mnras] {10.1093/mnras/stae1655}, \href {https://ui.adsabs.harvard.edu/abs/2024MNRAS.532.2584Y} {532, 2584}

\bibitem[\protect\citeauthoryear{{Zhu}, {Xue}, {Mao}, {Yang}  \& {Zhang}}{{Zhu} et~al.}{2025}]{ling2025}
{Zhu} L.,  {Xue} X.-X.,  {Mao} S.,  {Yang} C.,   {Zhang} L.,  2025, \mn@doi [\aap] {10.1051/0004-6361/202556036}, \href {https://ui.adsabs.harvard.edu/abs/2025A&A...703A..43Z} {703, A43}

\bibitem[\protect\citeauthoryear{{van de Ven}, {Hunter}, {Verolme}  \& {de Zeeuw}}{{van de Ven} et~al.}{2003}]{vandeven2003}
{van de Ven} G.,  {Hunter} C.,  {Verolme} E.~K.,   {de Zeeuw} P.~T.,  2003, \mn@doi [\mnras] {10.1046/j.1365-8711.2003.06501.x}, \href {https://ui.adsabs.harvard.edu/abs/2003MNRAS.342.1056V} {342, 1056}

\makeatother
\end{thebibliography}




\appendix


\bsp	
\label{lastpage}
\end{document}